\documentclass[preprintnumbers,amsmath,amssymb,floatfix,10pt,prd,onecolumn,superscriptaddress,nofootinbib]{revtex4}
\usepackage[colorlinks=true, pdfstartview=FitV, linkcolor=magenta, citecolor=red, urlcolor=blue]{hyperref}
\usepackage{epsfig,color}
\usepackage{xcolor}
\usepackage{epstopdf}
\usepackage{amssymb}
\usepackage{verbatim}
\usepackage{multirow}
\usepackage{latexsym}
\usepackage{epsfig}
\usepackage{epstopdf}
\usepackage{graphicx}
\usepackage{amssymb}
\usepackage{amsmath}
\usepackage{dcolumn}
\usepackage{bm}
\usepackage{color}
\usepackage{comment}

\usepackage{mathrsfs}
\usepackage{calrsfs}
\DeclareMathAlphabet{\pazocal}{OMS}{zplm}{m}{n}

\begin{document} \sloppy

\title{Traversable Wormholes in non-minimal Einstein-Yang-Mills Gravity: Geometry, Energy Conditions, and Gravitational Lensing}

\author{Jureeporn Yuennan}
\email{jureeporn\_yue@nstru.ac.th}
\affiliation{Faculty of Science and Technology, Nakhon Si Thammarat Rajabhat University, Nakhon Si Thammarat, 80280, Thailand}

\author{Allah Ditta}
\email{mradshahid01@gmail.com}
\affiliation{Department of Mathematics, School of Science, \\University of Management and Technology,  Lahore, 54000, Pakistan.}
\affiliation{Research Center of Astrophysics and Cosmology, Khazar University, Baku, AZ1096, 41 Mehseti Street, Azerbaijan}

\author{Thammarong Eadkhong}
\email{thammarong.ea@mail.wu.ac.th}
\affiliation{School of Science, Walailak University, Nakhon Si Thammarat, 80160, Thailand}

\author{Phongpichit Channuie}
\email{phongpichit.ch@mail.wu.ac.th}
\affiliation{School of Science, Walailak University, Nakhon Si Thammarat, 80160, Thailand}
\affiliation{College of Graduate Studies, Walailak University, Nakhon Si Thammarat, 80160, Thailand}

\begin{abstract}
This work presents a new class of static, spherically symmetric traversable wormhole solutions within the framework of non-minimal Einstein-Yang-Mills (EYM) gravity, where the SU(2) Yang-Mills field is purely magnetic. By adopting a constant redshift function and introducing a direct coupling between the Ricci scalar and the Yang-Mills field strength, we investigate the role of the non-minimal coupling constant $\xi$ and the magnetic charge $Q$ in shaping the wormhole geometry. Our analysis shows that for small values of $\xi$, the flare-out and throat conditions can be satisfied, allowing physically viable traversable wormholes without requiring externally introduced exotic matter. The Arnowitt-Deser-Misner (ADM) mass is evaluated, revealing that for $\xi < 0.01$ it grows monotonically with charge, whereas for $\xi \gtrsim 0.01$ it decreases with increasing charge, signaling a reduction in the total mass-energy of the system. An examination of the energy conditions indicates localized violations of the null and weak energy conditions at the throat, while the strong energy condition remains satisfied. Finally, the study of gravitational lensing confirms that the deflection angle of light is consistently positive, reflecting the overall attractive nature of the wormhole’s gravitational field. These results highlight the significant role of non-minimal gauge-gravity couplings in enabling traversable wormholes with distinct observational signatures.
\end{abstract}


\maketitle


\section{Introduction}

Wormholes have been proposed as formations in spacetime that operate as connectors between distant regions within the same universe or as linkages between wholly separate universes \cite{visser1995lorentzian}. Recent scientific scrutiny to these phenomena was inspired by the work of Morris and Thorne \cite{morris1988wormholes}, who proposed and investigated a particular type known as "traversable wormholes." Previous research investigations have examined wormhole solutions containing various critical curves, especially in structures where the effective potential is asymmetric around the throat of the wormhole \cite{morris1988wormholes, morris1988wormholes1,visser1995lorentzian,visser2003traversable,lobo2005phantom, blazquez2022einstein, capozziello2021traversable, blazquez2021traversable, berry2020thin, maldacena2021humanly, wielgus2020reflection, blazquez2021ellis, rueda2025traversable}. Recently, it has been conclusively proven that all traversable wormholes, regardless symmetric or asymmetric, have to show at least one critical curve \cite{xavier2024traversable}. This study found that although asymmetry affects the placement of these curves, symmetric wormholes invariably possess one precisely at the wormhole throat. Following on previous findings, this paper proposes  symmetric traversable wormhole models. A key aspect of our structure is the incorporation of a nonzero redshift function, $\Phi(r) \neq 0$. This element is crucial in influencing the spacetime geometry of the wormhole and defining observable phenomena. A vanishing redshift function helps mitigate radial tidal forces, but it also limits the intricacy of the effective potential, hence hindering the emergence of many unstable critical curves. Conversely, a nonzero $\Phi(r)$ generates radial energy gradients that facilitate the emergence of supplementary critical curves. These are essential for producing distinctive optical characteristics that may be observable. This decision is based on physically feasible conditions in which gravitational redshift influences the trajectories of photons and the characteristics of emitted radiation. The shape function $b(r)$, which essentially dictates the throat geometry, exerts limited influence on the existence of these curves, illustrating the significant relevance of the redshift function in this context.

In modern cosmology, the concept of wormholes has become a topic of significant theoretical interest due to their potential implications for the structure of spacetime. A wormhole is generally described as a hypothetical tunnel-like structure that connects two separate points in spacetime, which may lie within the same universe or span across different universes. This idea was originally introduced as a tool to explore the deeper implications of General Relativity (GR) \cite{morris1988wormholes2} . Several theoretical studies have reported encouraging indications within GR that support the plausibility of wormhole geometries \cite{rahaman2014possible,kuhfittig2014gravitational,lukmanova2016gravitational,li2014distinguishing,abe2010gravitational,toki2011astrometric}. More recent classifications distinguish wormholes into two broad types: static, which remain unchanging over time, and dynamic, which evolve with time \cite{jamil2013observational}.

A fundamental requirement for maintaining a traversable wormhole involves the presence of so-called exotic matter—substances or fields that violate the null energy condition (NEC) \cite{morris1988wormholes2,visser1996voting}. The violation of such energy conditions is essential because standard matter, which satisfies these conditions, does not allow for the stable geometry required by a wormhole, see Casimir traversable wormholes \cite{Garattini:2019ivd} and their extension/modifications \cite{Samart:2021tvl,Garattini:2025gfq,Jusufi:2020rpw,Garattini:2020kqb,Garattini:2021kca,Garattini:2024jkr,Sarkar:2025iiz}. This creates a challenge in GR, as its traditional framework does not naturally accommodate such matter distributions. However, this limitation can be addressed by altering the stress-energy tensor, which effectively allows for controlled violations of the NEC and enables wormhole solutions within GR. The foundational work in this area was presented by Einstein and Rosen in 1935, where they formulated a solution now known as the Einstein-Rosen bridge or Schwarzschild wormhole \cite{einstein1935particle}. Since then, extensive research has been conducted to derive wormhole solutions by incorporating a variety of exotic matter forms and theoretical constructs. These include scalar fields, quintom models, non-commutative geometry, and electromagnetic fields \cite{kim2001exact,rahaman2012searching,lobo2013new,capozziello2012wormholes,capozziello2008extended,capozziello2005reconciling,capozziello2011hydrostatic}. In each case, the modified matter content was carefully analyzed in the context of the energy conditions and gravitational field equations. Interestingly, some investigations have shown that under certain conditions, wormholes may exist without requiring exotic matter, challenging the long-held belief that such matter is always necessary \cite{kanti2011wormholes,kanti2012stable}. These results are particularly valuable as they offer alternative paths to constructing physically realistic wormholes within modified gravity frameworks or under specific configurations of the matter content.

Wormholes may also be sustained by electromagnetic fields and non-Abelian gauge fields, contingent upon these fields satisfying specific exotic energy criteria. This situation is feasible when nonlinear electrodynamics is considered \cite{arellano2006evolving, arellano2006non}, or when gravity is non-minimally coupled with vector fields, such as the non-Abelian Yang-Mills field or the Maxwell field. The non-minimal Einstein-Maxwell paradigm has been intensively explored in both its linear~\cite{faraoni1998conformal,hehl2001does} and nonlinear~\cite{balakin2005non} versions. Additionally, more recently, a new class of WH solutions within the framework of Einstein-Euler–Heisenberg (EEH) nonlinear electrodynamics has been investigated in Ref.\cite{Channuie:2025xlw}. Within the framework of non-minimal Einstein-Yang-Mills theory, two recognized methodologies exist for deriving the fundamental equations. One method entails a dimensional reduction of the Gauss-Bonnet action~\cite{muller1988modification}, yielding a model defined by a singular coupling constant. The other way extends the non-minimal nonlinear Einstein-Maxwell theory to a non-Abelian setting~\cite{balakin2005non,balakin2007nonminimal}.

In the framework of non-minimal Einstein--Yang--Mills (EYM) gravity, the investigation of light deflection offers critical insights into the underlying spacetime geometry and the influence of gauge fields on photon trajectories. The non-minimal coupling between the Yang--Mills field and the gravitational sector modifies the curvature of spacetime, which directly impacts how light rays bend as they pass through or near the wormhole throat \cite{Balakin2005, Dyadichev2002}. The deflection angle, in this context, depends on both the form of the metric functions and the strength of the non-minimal coupling. A key feature of wormhole geometries in EYM gravity is the absence of an event horizon, which permits the photons to probe regions near or even through the throat. For physically admissible configurations, one typically finds a positive deflection angle, indicating that the gravitational field retains an overall attractive character in the vicinity of the wormhole \cite{Bhattacharya2010}. This positive bending confirms the photon’s path is curved toward the wormhole, analogous to standard gravitational lensing but distinct due to the unique topological structure and the role of the Yang--Mills field. The behavior of the deflection angle also serves as a diagnostic tool to differentiate wormholes from other compact objects. In particular, the magnitude of positive deflection may reveal constraints on the shape function and the non-minimal coupling parameters \cite{Mishra2018}. Hence, by examining light bending in such settings, one gains valuable information about the physical viability of wormhole solutions and the interplay between matter fields and spacetime curvature in modified gravity theories.

This paper is organized as follows: Section \ref{sec2}: Presents the theoretical foundation, detailing the non-minimally-coupled EYM action and the general metric for static, spherically symmetric wormholes. In the same section, we derive the wormhole solutions for a constant redshift function model. Section \ref{sec3}: we present the embedding diagrams employed to illustrate the non-minimal EYM wormhole. In Section \ref{sec4}, we calculate the ADM mass and explores its dependence on model parameters. In Section \ref{sec5}, we analyze the energy conditions for a model of the constant redshift function, $\Phi=const$, exploring their implications for the wormhole's traversability and stability. Section \ref{sec6}: Examines the gravitational lensing effects, emphasizing the influence of the non-minimal coupling constant ($\xi$) and the Yang–Mills magnetic charge (Q). We conclude our findings in the last section.

\section{Action and Field Equations}\label{sec2}

We consider the action for non-minimal Einstein--Yang--Mills (EYM) theory \cite{Lutfuoglu:2025ljm}
\begin{equation}
S = \int d^4x \sqrt{-g} \left[ \frac{R}{16\pi G} - \frac{1}{4} F^a_{\mu\nu} F^{a\mu\nu} + \frac{\xi}{4} R F^a_{\mu\nu} F^{a\mu\nu} \right],
\end{equation}
where $R$ is the Ricci scalar, $F^a_{\mu\nu}$ is the Yang--Mills field strength tensor of the SU(2) Yang--Mills gauge field, and $\xi$ is the non-minimal coupling constant. The first term in the integrand corresponds to the standard Einstein--Hilbert action, governing the dynamics of spacetime geometry. The second term is the conventional Yang–Mills action, describing the dynamics of the gauge fields. The third term introduces a direct coupling between the Ricci scalar and the Yang–Mills field strength, leading to modifications in the interaction between gravity and gauge fields. This can significantly alter the energy-momentum tensor, possibly allowing violations of the null energy condition (NEC) without introducing exotic matter by hand -- a key requirement for sustaining traversable wormholes. We are putting the speed of light and the gravitational constant equal
to one, $c=1$, $G=1$, so that $\frac{8\pi G}{c^4}=8 \pi$. The
SU(2) Yang-Mills field is described by a triplet of
vector potentials $A^{a}_\mu$, where the group index $(a)$ runs
over three values. The Yang-Mills field
components $F^{a}_{\mu\nu}$ are
connected to the potentials $A^{a}_\mu$ by the relation:
\begin{equation}
F^{a}_{\mu\nu} = \nabla_\mu A^{a}_\nu - \nabla_\nu A^{a}_\mu +
f^{a}_{b\,c} A^{b}_\mu A^{c}_\nu \,. \label{Fmn}
\end{equation}
Here $\nabla _\mu$ is a  covariant spacetime derivative, and
the symbols $f^{a}_{b\,c}=\epsilon^{a}_{b\,c}$ denote the real structure
constants of the gauge group $SU(2)$. Defining $\mathcal{F} = F^a_{\mu\nu} F^{a\mu\nu}$ and varying the action with respect to the metric $g^{\mu\nu}$ yields the modified Einstein equations:
\begin{equation}
\frac{1}{8\pi G} G_{\mu\nu} =T^{({\rm eff.})}_{\mu\nu}\equiv T^{(\text{YM})}_{\mu\nu} + T^{(\text{non-min})}_{\mu\nu},
\end{equation}
where
\begin{align}
T^{\text{YM}}_{\mu\nu} &= F^a_{\mu\lambda} F_\nu^{a\ \lambda} - \frac{1}{4} g_{\mu\nu} \mathcal{F}, \\
T^{\text{non-min}}_{\mu\nu} &= \frac{\xi}{2} \left[ \Big(R_{\mu\nu}-\frac{1}{2}g_{\mu\nu}R\Big)\mathcal{F}+\nabla_{\mu}\nabla_{\mu}\mathcal{F}-g_{\mu\nu}\Box\mathcal{F}\right]\,,
\end{align}
where
\begin{align}
\Box\mathcal{F}\equiv \nabla^{\mu}\nabla_{\mu}\mathcal{F}=\frac{1}{\sqrt{-g}}\partial_{\mu}\Big(\sqrt{-g}g_{\mu\nu}\partial_{\nu}\mathcal{F}\Big)\,.
\end{align}
We now consider a static, spherically symmetric wormhole metric of the form:
\begin{equation}
ds^2 = -e^{2\Phi(r)} dt^2 + \left(1 - \frac{b(r)}{r} \right)^{-1} dr^2 + r^2(d\theta^2 + \sin^2\theta\, d\phi^2),\label{line}
\end{equation}
where the functions $\Phi(r)$ and $b(r)$ denote the redshift and shape functions, respectively, characterizing a static and spherically symmetric wormhole geometry. These functions are defined over the radial domain $r \in [r_0, +\infty)$ and can take arbitrary forms within this range \cite{morris1988wormholes,morris1988wormholes1}. A key feature of a traversable wormhole is the flare-out condition, expressed as $(b - b'r)/b^2 > 0$, which must be satisfied in conjunction with the requirement $1 - b(r)/r > 0$. At the wormhole throat, the condition $b(r_0) = r_0$ holds, and the additional constraint $b'(r_0) < 1$ is necessary to support a wormhole solution. Moreover, to ensure traversability and avoid the presence of horizons defined as surfaces where $e^{2\Phi(r)} \to 0$ -- the redshift function $\Phi(r)$ must remain finite throughout the entire spacetime. With the help of the line element (\ref{line}),
we can write the effective field equations (EFEs) in an orthonormal reference frame, leading to the
following set of equations%
\begin{eqnarray}
\frac{b^{\prime }\left( r\right) }{r^{2}}&=&8\pi \rho^{\rm {eff.}} \left( r\right) ,
\label{rho}\\
\frac{2}{r}\left( 1-\frac{b\left( r\right) }{r}\right) \Phi \!^{\prime
}\left( r\right) -\frac{b\left( r\right) }{r^{3}}&=&8\pi p^{\rm {eff.}}_{r}\left(
r\right) ,  \label{pr}\\
\left( 1-\frac{b\left( r\right) }{r}\right) \left[ \Phi ^{\prime \prime
}\!\left( r\right) +\Phi \!^{\prime }\left( r\right) \left( \Phi ^{\prime
}\!\left( r\right) +\frac{1}{r}\right) \right] -\frac{b^{\prime }\left(
r\right) r-b\left( r\right) }{2r^{2}}\left( \Phi \!^{\prime }\left( r\right)
+\frac{1}{r}\right) &=&8\pi p^{\rm {eff.}}_{t}(r), \label{pt}
\end{eqnarray}%
in which $\rho^{\rm {eff.}} \left( r\right) $ is the (effective) energy density, $p^{\rm {eff.}}_{r}\left(
r\right) $ is the (effective) radial pressure, and $p^{\rm {eff.}}_{t}\left( r\right) $ is the (effective) lateral pressure and $G=1$. The wormhole throat is located at $r = r_0$ such that $b(r_0) = r_0$.
The d'Alembertian (curved-space Laplacian) acting on a scalar field \(\mathcal{F}(r)\) is defined as:
\[
\Box \mathcal{F} = \nabla^\mu \nabla_\mu \mathcal{F}
= \frac{1}{\sqrt{-g}} \partial_\mu \left( \sqrt{-g} \, g^{\mu\nu} \partial_\nu \mathcal{F} \right)\,.
\]
Neglecting the angular \(\theta\)-dependence (as it cancels), we obtain:
\begin{eqnarray}
\Box \mathcal{F} = \frac{1}{r^2} \frac{d}{dr} \left[
r^2 \left(1 - \frac{b(r)}{r} \right) \frac{d\mathcal{F}}{dr}
\right] + \Phi'(r) \left(1 - \frac{b(r)}{r} \right) \frac{d\mathcal{F}}{dr}\,,
\end{eqnarray}
with
\[
\sqrt{-g} = e^{\Phi(r)} r^2 \sin\theta \left(1 - \frac{b(r)}{r} \right)^{-1/2}
\quad \text{and} \quad
g^{rr} = 1 - \frac{b(r)}{r}\,.
\]
The Ricci scalar $R$ for this metric is given by:
\begin{eqnarray}
R &=& \frac{b'(r)}{r^2} \Big( 2+ r\Phi'(r) \Big) -\frac{2}{r}\Big(2\Phi'(r)+r\Phi'^{2}(r)+r\Phi''(r)\Big)\nonumber\\&+&\frac{b(r)}{r^{2}}\Big(3\Phi'(r)+2r\Phi'^{2}(r)+2r\Phi''(r)\Big).
\end{eqnarray}
Notice that in the special case of a constant redshift function ($\Phi'(r) = 0$), the scalar curvature simplifies to $R = \tfrac{2b'(r)}{r^2}$. While a general nonzero redshift function affects optical properties, in this work we focus on the constant $\Phi$ case to allow analytical progress and direct comparison with existing wormhole solutions. The constant $\Phi$ model also isolates the geometric effects of the non-minimal coupling from those induced by strong gravitational redshift variations. The Wu--Yang ansatz for a purely magnetic SU(2) gauge field is given by \cite{Balakin:2015gpq}
\begin{equation}
A^a_0 = 0, \quad A^a_r = 0, \quad A^a_\theta = \delta^a_{(\varphi)}, \quad A^a_\varphi = -Q \sin\theta\, \delta^a_{(\theta)},
\end{equation}
with $Q$ being a non-zero integer. From this ansatz, the only non-zero component of the field strength tensor is:
\begin{equation}
F^a_{\theta\varphi} = \partial_\theta A^a_\varphi - \partial_\varphi A^a_\theta + \epsilon^a{}_{bc} A^b_\theta A^c_\varphi = Q \sin\theta\, \delta^a_{r}.
\end{equation}
In the given metric, the relevant inverse components are:
\[
g^{\theta\theta} = \frac{1}{r^2}, \quad g^{\varphi\varphi} = \frac{1}{r^2 \sin^2\theta}.
\]
Raising indices yield
\begin{equation}
F^{a\,\theta\varphi} = g^{\theta\theta} g^{\varphi\varphi} F^a_{\theta\varphi} = \frac{Q}{r^4 \sin\theta} \delta^a_{r}.
\end{equation}
Hence, the contraction gives for $\mathcal{F}$
\begin{align}
\mathcal{F} &= F^a_{\mu\nu} F^{a\,\mu\nu} = 2 F^a_{\theta\varphi} F^{a\,\theta\varphi} \\
&= 2 \left( Q \sin\theta\, \delta^a_{r} \cdot \frac{Q}{r^4 \sin\theta} \delta^a_{r} \right) = \frac{2Q^2}{r^4}.
\end{align}
Since the only non-zero components of $T^{\text{YM}}_{\mu\nu}$ are diagonal, we obtain
\begin{equation}
T^{({\rm YM})\,t}_{t}=T^{({\rm YM})\,r}_{r}=-\frac{Q^{2}}{r^{4}},\,T^{({\rm YM})\,\theta}_{\theta}=T^{({\rm YM})\,\phi}_{\phi}=\frac{Q^{2}}{r^{4}}\,.
\end{equation}
Therefore, the total effective Stress Energy Tensor (SET) becomes
\begin{equation}
T^{({\rm eff.})}_{\mu\nu}=-\frac{Q^{2}}{r^{4}}{\rm diag.}(1,1,-1,-1)+T^{(\text{non-min})}_{\mu\nu}\,,
\end{equation}
where
\begin{eqnarray}
\rho^{(\text{non-min})}(r)&=&-\frac{4\xi  Q^2}{r^6}\left[2 \left(\frac{r b'(r)-b(r)}{r}+3 \left(\frac{b(r)}{r}+1\right)\right)+b'(r)\right]\,,\\
p^{(\text{non-min})}_{r}(r)&=&\frac{1}{2} \xi  \Bigg[-\frac{2 Q^2 b'(r)}{r^6}-\frac{4 Q^2}{r^7}\left(-\frac{8 Q^2}{r^6}\left(\frac{r b'(r)-b(r)}{r}+3 \left(\frac{b(r)}{r}+1\right)\right)-r b'(r)+b(r)\right)\nonumber\\&&+\frac{40 Q^2}{r^6}\left(1-\frac{b(r)}{r}\right)\Bigg]\,,\\
p^{(\text{non-min})}_{t}(r)&=&\frac{1}{2} \xi  \Bigg[\frac{2 Q^2}{r^7}\left(b(r)-r b'(r)\right)-\frac{8 Q^2}{r^6}\left(\frac{r b'(r)-b(r)}{r}+3 \left(\frac{b(r)}{r}+1\right)\right)\Bigg]\,.
\end{eqnarray}
In the following, we focus only on $\Phi'(r) = cont.$. Therefore, the set of equations given above reduces to:
\begin{eqnarray}
\frac{b^{\prime }\left( r\right) }{r^{2}}&=&8\pi \rho^{\rm {eff.}}(r) ,
\label{rho1}\\
-\frac{b\left( r\right) }{r^{3}}&=&8\pi p^{\rm {eff.}}_{r}(r),  \label{pr1}\\
-\frac{b^{\prime }\left(
r\right) r-b\left( r\right) }{2r^{3}} &=&8\pi p^{\rm {eff.}}_{t}(r), \label{pt1}
\end{eqnarray}%
and
\[
\Box \mathcal{F} = \frac{1}{r^2} \frac{d}{dr} \left[
r^2 \left(1 - \frac{b(r)}{r} \right) \frac{d\mathcal{F}}{dr}
\right]\,,
\]
\begin{widetext}
\begin{eqnarray}
b(r)&:=&\frac{r_{0}^{7/3} \sqrt[3]{96 \pi  \xi  Q^2+r^4}}{r^{4/3} \sqrt[3]{96 \pi  \xi  Q^2+r_{0}^4}}+\nonumber\\&&+\frac{1}{34944 \pi^{4/3} \xi  r^{4/3} \sqrt[3]{96 \pi  \xi  Q^2+r_{0}^4}}\Bigg[\sqrt[3]{2} 3^{2/3} \Bigg(r_{0}^{7/3} \sqrt[3]{96 \pi  \xi  Q^2+r^4} \sqrt[3]{\frac{r_{0}^4}{\xi  Q^2}+96 \pi } \nonumber\\&&\Bigg(2496 \pi  \xi  \, _2F_1\left(\frac{7}{12},\frac{4}{3};\frac{19}{12};-\frac{r_{0}^4}{96 \pi  Q^2 \xi }\right)-7 r_{0}^2 \, _2F_1\left(\frac{13}{12},\frac{4}{3};\frac{25}{12};-\frac{r_{0}^4}{96 \pi  Q^2 \xi }\right)\Bigg)\nonumber\\&&-2496 \pi  \xi  r^{7/3} \sqrt[3]{\frac{r^4}{\xi  Q^2}+96 \pi } \sqrt[3]{96 \pi  \xi  Q^2+r_{0}^4} \, _2F_1\left(\frac{7}{12},\frac{4}{3};\frac{19}{12};-\frac{r^4}{96 \pi  Q^2 \xi }\right)\nonumber\\&&+7 r^{13/3} \sqrt[3]{\frac{r^4}{\xi  Q^2}+96 \pi } \sqrt[3]{96 \pi  \xi  Q^2+r_{0}^4} \, _2F_1\left(\frac{13}{12},\frac{4}{3};\frac{25}{12};-\frac{r^4}{96 \pi  Q^2 \xi }\right)\Bigg)\Bigg]\,,\label{brf}
\end{eqnarray}
\end{widetext}
where $_{2}F_{1} (a,b; c; z)$ denotes the hypergeometric function. Here we have chosen $\xi>0, Q>0$, and $r_{0}>0$ such that the argument of the hypergeometric function remains within the real-valued domain. Numerical evaluations were performed with these constraints to avoid complex-valued solutions.
\begin{figure}
\includegraphics[width=8 cm]{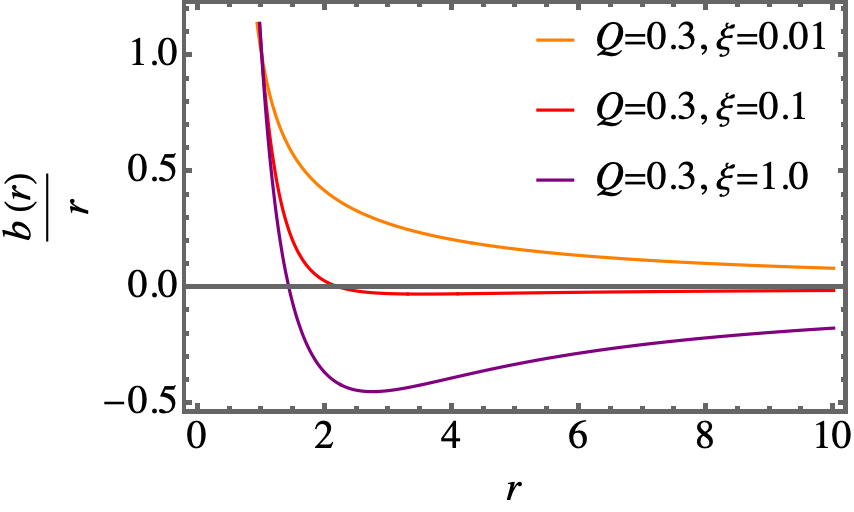}
\includegraphics[width=8 cm]{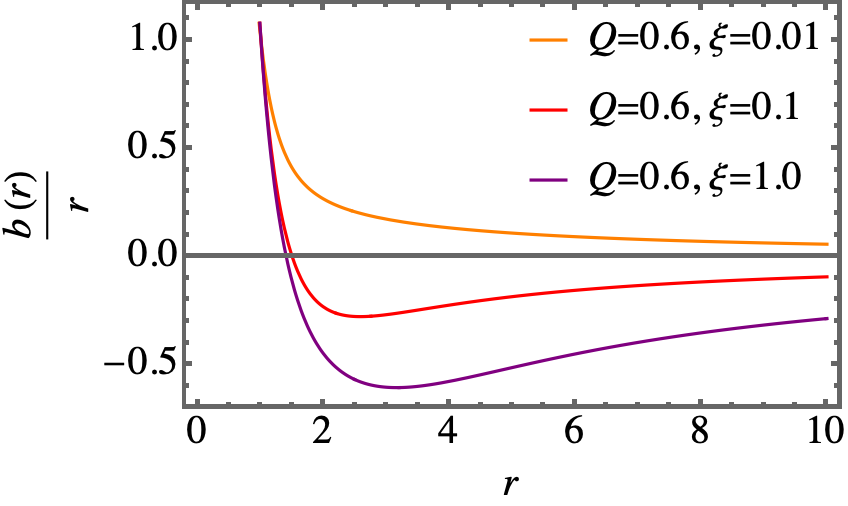}
\includegraphics[width=8 cm]{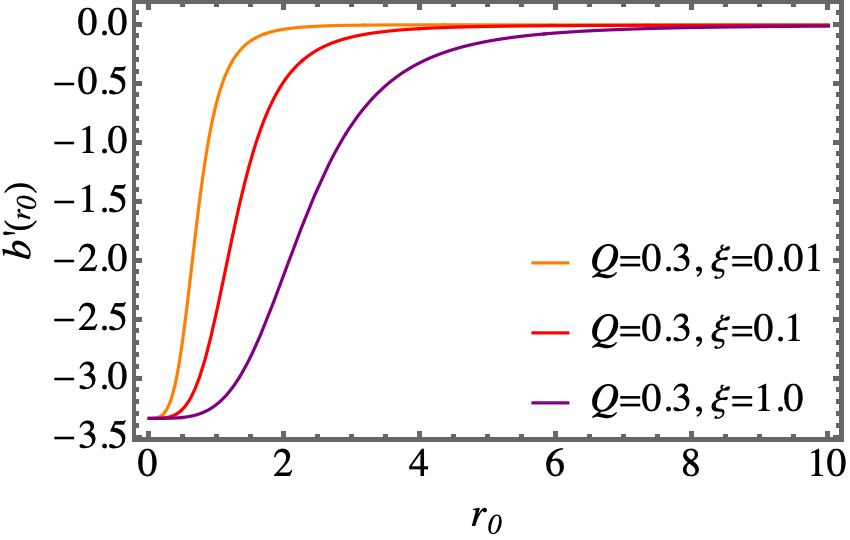}
\includegraphics[width=8 cm]{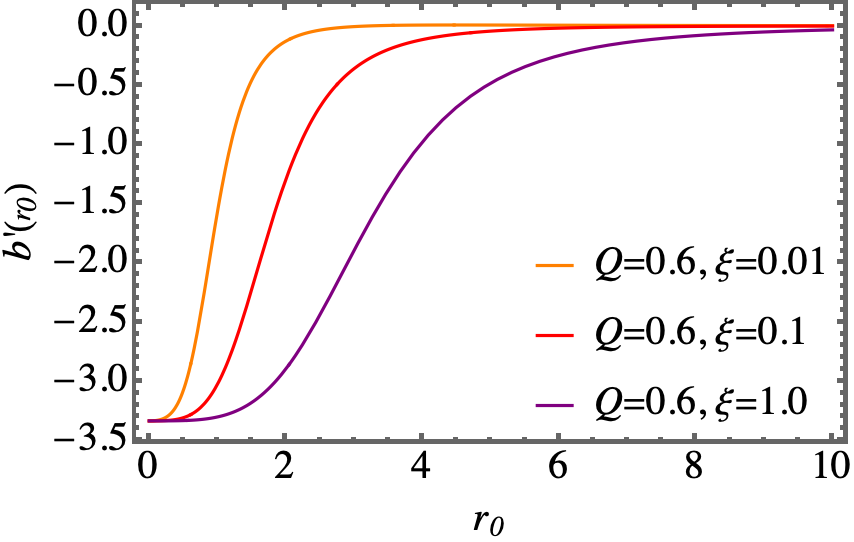}
\includegraphics[width=8 cm]{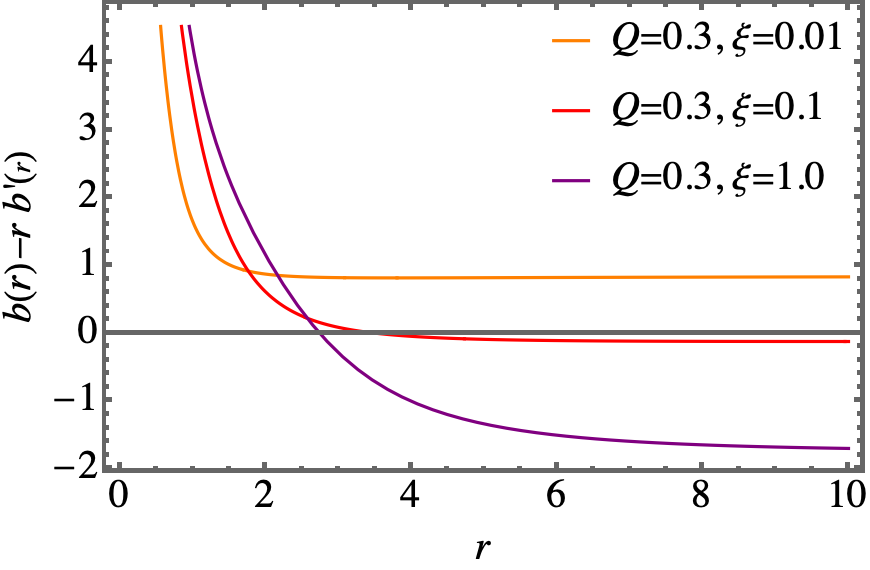}
\includegraphics[width=8 cm]{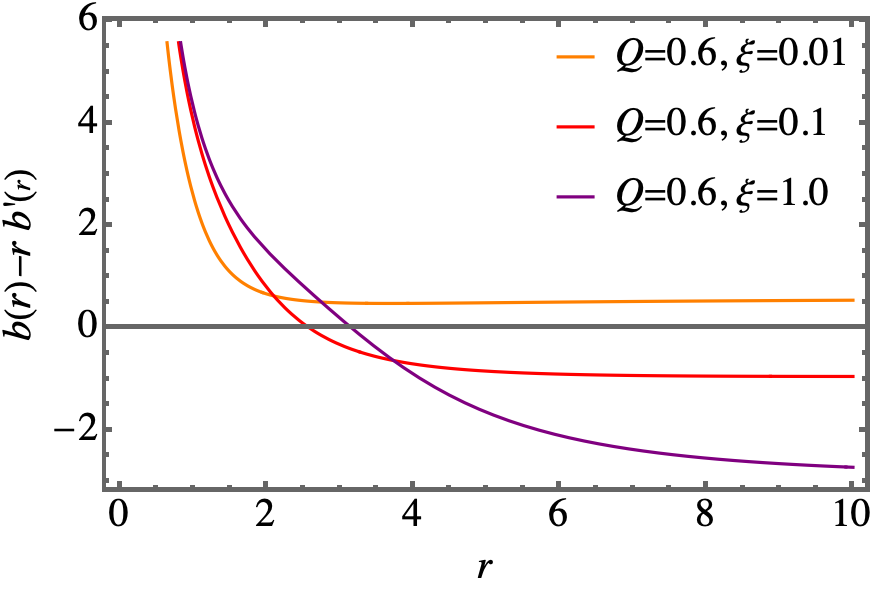}
\caption{Upper panels: the behavior of the ratio $b(r)/r$ as a function of $r$ for various values of the coupling constant $\xi$, plotted for $Q = 0.3$ (left) and $Q = 0.6$ (right). Center panels: we displayed the additional constraint $b'(r_0) < 1$ for $Q = 0.3$ (left) and $Q = 0.6$ (right), and Lower panels: the plots show the flare-out condition, expressed as $(b - b'r)/b^2 > 0$ for $Q = 0.3$ (left) and $Q = 0.6$ (right).}\label{brpr}
\end{figure}

The upper panels of Fig.~\ref{brpr} presents the radial profiles of the dimensionless ratio $b(r)/r$ for two values of the Yang--Mills magnetic charge: $Q = 0.3$ (left panel) and $Q = 0.6$ (right panel), each examined with three different non-minimal coupling strengths: $\xi = 0.01$, $0.1$, and $1.0$. These plots help assess the wormhole geometry's compliance with traversability conditions. For both values of $Q$, it is observed that for small $\xi$ (e.g., $\xi = 0.01$), the ratio $b(r)/r$ remains positive and decreases monotonically with increasing $r$, indicating a standard flaring-out geometry required for traversability. However, as the coupling constant $\xi$ increases, the function $b(r)/r$ begins to dip below zero near the throat and displays non-monotonic behavior. In particular, for $\xi = 1.0$, the ratio becomes negative over a significant range, especially in the vicinity of the throat. This implies a violation of the flare-out condition, i.e., $1 - b(r)/r > 0$ is not upheld, which could signal either non-traversability or a breakdown of the classical wormhole structure in this regime. However, such behavior also hints at possible quantum-modified geometries or exotic topologies arising from strong non-minimal interactions. The impact of the Yang--Mills charge $Q$ is also evident. Increasing $Q$ leads to a more pronounced suppression of $b(r)/r$ in the region near the throat, especially for larger $\xi$. This indicates a coupling between the gauge field strength and the geometric deformation introduced by the non-minimal interaction. The combined effects of $Q$ and $\xi$ suggest that the shape of the wormhole can be finely tuned by adjusting these parameters, allowing the potential realization of physically admissible traversable wormhole geometries in specific regions of parameter space.

The center panels of Fig.~\ref{brpr} display the variation of the shape function derivative at the throat, $b'(r_0)$, as a function of $r_0$, for different values of the non-minimal coupling constant $\xi$ and two fixed values of the Yang--Mills magnetic charge $Q = 0.3$ (left panel) and $Q = 0.6$ (right panel). These plots are crucial for assessing the flare-out condition $b'(r_0) < 1$ and ensuring wormhole traversability. For small throat radii ($r_0 < 2$), all cases exhibit strongly negative values of $b'(r_0)$, confirming that the flare-out condition is well satisfied. As $r_0$ increases, the values of $b'(r_0)$ asymptotically approach zero from below, indicating a weakening of the flaring behavior with increasing throat size. Notably, for larger $\xi$ values, the magnitude of $b'(r_0)$ is significantly more negative at small $r_0$, suggesting that stronger non-minimal coupling enhances the flaring at small scales. Comparing both panels, the effect of increasing the Yang--Mills charge from $Q = 0.3$ to $Q = 0.6$ appears to slightly shift the curves downward, indicating that larger $Q$ values intensify the negativity of $b'(r_0)$ for small throats. However, this effect is less prominent than that of the coupling constant $\xi$. The results imply that for appropriately chosen values of $\xi$ and $Q$, the shape function derivative satisfies the necessary conditions for wormhole solutions across a broad range of $r_0$. This supports the existence of stable wormhole throats within the non-minimal Einstein--Yang--Mills framework and emphasizes the role of the non-minimal interaction in modulating the geometry near the throat.

The lowest panels of Fig.~\ref{brpr} illustrate the behavior of the flare-out condition, characterized by the quantity \( b(r) - r b'(r) \), for two fixed values of the Yang--Mills magnetic charge: \( Q = 0.3 \) (left panel) and \( Q = 0.6 \) (right panel). This function serves as a direct diagnostic for assessing the flare-out condition \( (b(r) - r b'(r))/b^2 > 0 \), which must be satisfied near the wormhole throat to ensure the geometry expands outward, thereby allowing traversability. Since the denominator \( b^2 \) is always positive, the sign of \( b(r) - r b'(r) \) alone determines whether the flare-out condition is met. In both panels, the curves correspond to different values of the non-minimal coupling parameter \( \xi \). When \( \xi = 0.01 \), the function remains positive across the entire radial domain, indicating that the flare-out condition is robustly satisfied throughout the wormhole geometry. As \( \xi \) increases to 0.1 and 1.0, however, the function begins to dip and eventually crosses the zero line. This transition signals the onset of a region where the flare-out condition is violated. The point at which this violation begins shifts to smaller radii as \( \xi \) increases, demonstrating that stronger non-minimal coupling tends to suppress the geometric flaring behavior required for a traversable wormhole. Furthermore, increasing the Yang--Mills charge from \( Q = 0.3 \) to \( Q = 0.6 \) accentuates this effect. For the same values of \( \xi \), the dip in \( b(r) - r b'(r) \) becomes deeper and extends over a broader range of \( r \) in the right panel. This behavior underscores the cooperative influence of \( \xi \) and \( Q \) in governing the satisfaction of the flare-out condition. In particular, large values of both parameters lead to a pronounced suppression of the condition, potentially inhibiting the formation of viable wormhole solutions. Overall, the figures suggest that while small values of the non-minimal coupling permit the flare-out condition to be naturally fulfilled, higher values of \( \xi \) and \( Q \) must be carefully constrained. Within suitable parameter ranges, the wormhole geometry can still satisfy the flare-out requirement without the explicit need for exotic matter. This finding highlights the geometrical plausibility of wormhole solutions in the context of the non-minimal Einstein--Yang--Mills framework.

\section{Embedding Diagram}\label{sec3}
This section presents the embedding diagrams employed to illustrate the EEH wormhole by focusing on an equatorial slice at $\theta = \pi/2$ and a fixed time, where $t$ remains constant. The corresponding spatial metric on this slice is given by
\begin{equation}
ds^2 = \frac{dr^2}{1 - \frac{b(r)}{r}} + r^2 d\phi^2. \label{emb}
\end{equation}
To visualize this geometry, we embed the metric in Eq.~(\ref{emb}) into a three-dimensional Euclidean space, which in cylindrical coordinates is described by
\begin{equation}
ds^2 = dz^2 + dr^2 + r^2 d\phi^2.
\end{equation}
By comparing the two metrics, we derive the embedding function as
\begin{equation}
\frac{dz}{dr} = \pm \sqrt{\frac{r}{r - b(r)} - 1}.
\end{equation}
The shape function $b(r)$ is specified by Eq.(\ref{brf}). Since this expression does not yield to analytical integration, we rely on numerical techniques to generate the embedding diagrams shown in Fig.\ref{f8}. These plots demonstrate the effect of the quantum correction parameter $\alpha$ on the wormhole’s spatial geometry.
\begin{figure}
\includegraphics[width=8.0 cm]{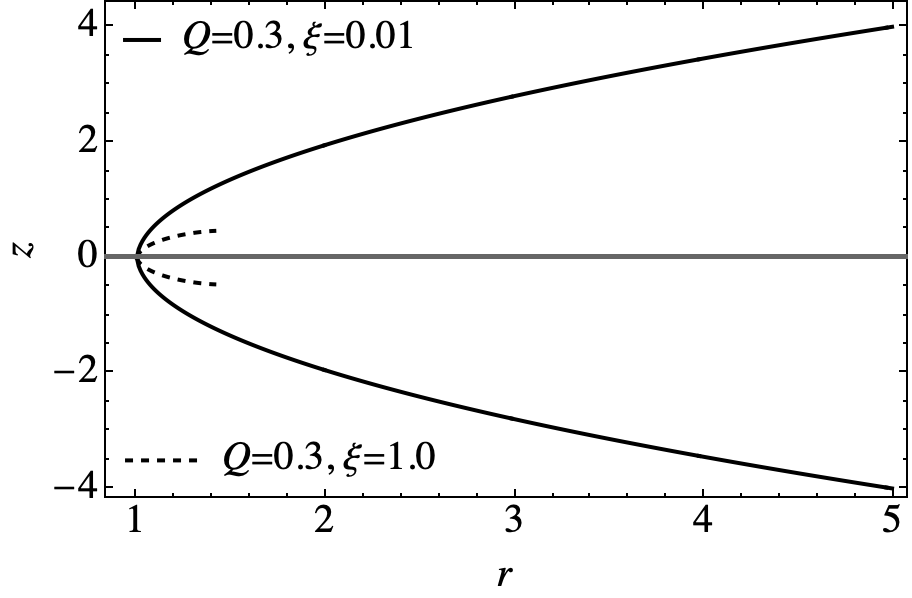}
\includegraphics[width=8.0 cm]{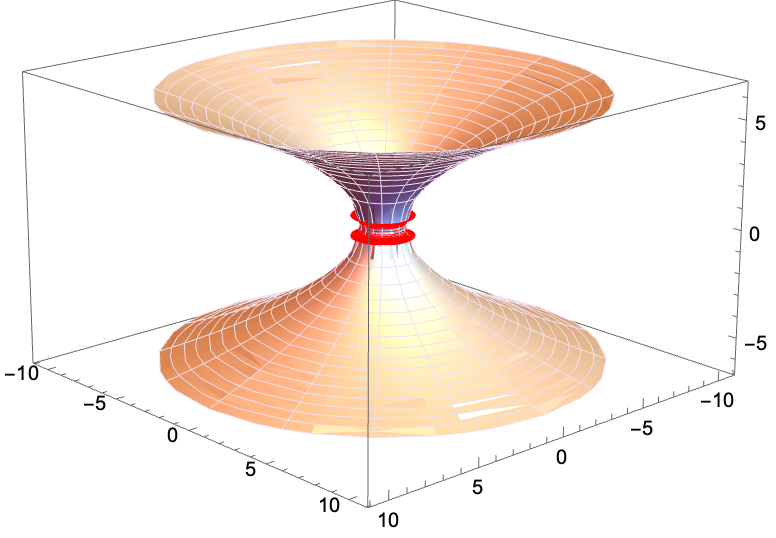}
\caption{Left: the 2D embedding diagram of the EEH wormhole. Here we have used $r_0=1,\,q=0.3$, and $\xi=0.01$ (solid line) and $\xi=0.1$ (solid line); Right: the 3D embedding diagram of the EEH wormhole using the same set of parameters.}\label{Embed}
\end{figure}

\section{ADM mass for non-minimal EYM wormholes}\label{sec4}
\begin{figure}
\includegraphics[width=9.0 cm]{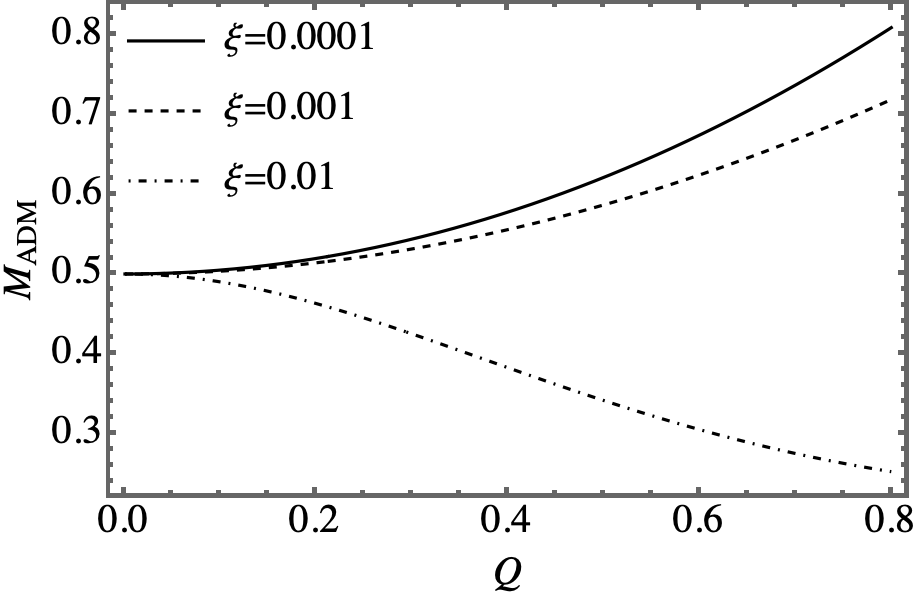}
\caption{The ADM mass, $M_{\rm ADM}$, is shown as a function of the black hole charge $Q$ for three values of the model parameter $\xi$: $0.0001$ (solid line), $0.001$ (dashed line), and $0.01$ (dash-dotted line). The figure demonstrates that $M_{\rm ADM}$ increases monotonically with $Q$ for $\xi < 0.01$. In contrast, for $\xi \gtrsim 0.01$, $M_{\rm ADM}$ decreases monotonically as $Q$ grows, indicating that the total mass-energy of the system diminishes with increasing electric charge.}\label{fm}
\end{figure}

We now proceed to evaluate the Arnowitt–Deser–Misner (ADM) mass associated with the non-minimal EYM wormhole. The ADM mass is a concept in general relativity that represents the total mass-energy content of a gravitational system as measured by an observer located at spatial infinity. The analysis begins with an asymptotically flat spacetime characterized by the metric
\begin{eqnarray}
ds^2_{\Sigma} = \psi(r) dr^2+ r^2 \chi(r) \left( d\theta^2 + \sin^2\theta , d\phi^2 \right),
\end{eqnarray}
where the metric functions are given by
\begin{equation}
\psi(r) = \frac{1}{1 - \frac{b(r)}{r}}, \quad \text{and} \quad \chi(r) = 1.
\end{equation}
To determine the ADM mass, we use the following expression (see \cite{Shaikh:2018kfv}):
\begin{equation}\label{for}
M_{\rm ADM} = \lim_{r \to \infty} \frac{1}{2} \left[-r^2 \chi' + r (\psi - \chi) \right].
\end{equation}
Substituting the relevant expressions and taking the limit yields the ADM mass as
\begin{eqnarray}
M_{\rm ADM}&=&\frac{1}{86016 \pi^{19/12} \xi r_{0}^{11/3} \Gamma \left(-\frac{2}{3}\right) \Gamma \left(\frac{1}{3}\right) \left(96 \pi  \xi  Q^2+r_{0}^4\right)^{7/3}}\Bigg[2016\ 3^{2/3} \pi ^{5/4} Q^2 r_{0}^4 \xi  \Big(35 r_{0}^8\nonumber\\&&+5312 \pi  Q^2 \xi  r_{0}^4+23552 \pi ^2 Q^4 \xi ^2\Big) \Gamma \left(\frac{1}{3}\right) \Gamma \left(-\frac{2}{3}\right) \, _2F_1\left(-\frac{2}{3},\frac{1}{12};\frac{13}{12};-\frac{r_{0}^4}{96 \pi  Q^2 \xi }\right) \nonumber\\&&\sqrt[3]{\frac{2 r_{0}^4}{Q^2 \xi }+192 \pi }+672\ 3^{2/3} \pi ^{5/4} Q^2 \xi  \Big(-5 r_{0}^{12}-1184 \pi  Q^2 \xi  r_{0}^8-56320 \pi ^2 Q^4 \xi ^2 r_{0}^4\nonumber\\&&+1081344 \pi ^3 Q^6 \xi ^3\Big) \Gamma \left(\frac{1}{3}\right) \Gamma \left(-\frac{2}{3}\right) \, _2F_1\left(-\frac{11}{12},\frac{1}{3};\frac{1}{12};-\frac{r_{0}^4}{96 \pi  Q^2 \xi }\right) \sqrt[3]{\frac{2 r_{0}^4}{Q^2 \xi }+192 \pi }\nonumber\\&&+\sqrt[4]{\pi } \Bigg(1105920\ 3^{2/3} \pi ^2 Q^2 r_{0}^6 \xi ^2 \left(r_{0}^4+96 \pi  Q^2 \xi \right) \Gamma \left(\frac{1}{3}\right) \Gamma \left(-\frac{2}{3}\right) \, _2F_1\left(-\frac{2}{3},\frac{7}{12};\frac{19}{12};-\frac{r_{0}^4}{96 \pi  Q^2 \xi }\right)\nonumber\\&& \sqrt[3]{\frac{2 r_{0}^4}{Q^2 \xi }+192 \pi }-7\ 3^{2/3} r_{0}^4 \Big(-15 r_{0}^{12}+5792 \pi  Q^2 \xi r_{0}^8+1264640 \pi ^2 Q^4 \xi ^2 r_{0}^4\nonumber\\&&+12812288 \pi ^3 Q^6 \xi^3\Big) \Gamma \left(-\frac{2}{3}\right) \Gamma \left(\frac{1}{3}\right) \, _2F_1\left(\frac{1}{12},\frac{1}{3};\frac{13}{12};-\frac{r_{0}^4}{96 \pi  Q^2 \xi }\right) \sqrt[3]{\frac{2 r_{0}^4}{Q^2 \xi }+192 \pi }\nonumber
\end{eqnarray}
\begin{eqnarray}
\nonumber\\&&-16 \sqrt[3]{\pi } \left(r_{0}^4+96 \pi  Q^2 \xi \right) \Bigg(-258048 \pi ^2 Q^2 r_{0}^6 \xi ^2 \Gamma \left(-\frac{2}{3}\right) \Gamma \left(\frac{1}{3}\right)\nonumber\\&&-448000 \pi ^2 Q^4 r_{0}^4 \xi ^2 \Gamma \left(-\frac{2}{3}\right) \Gamma \left(\frac{1}{3}\right)-2688 \pi  r_{0}^{10} \xi  \Gamma \left(-\frac{2}{3}\right) \Gamma \left(\frac{1}{3}\right)\nonumber\\&&+35 r_{0}^{12} \Gamma \left(\frac{1}{3}\right) \Gamma \left(-\frac{2}{3}\right)+5677056 \pi^3 Q^6 \xi ^3 \Gamma \left(\frac{1}{3}\right) \Gamma \left(-\frac{2}{3}\right)\nonumber\\&&+1064 \pi  Q^2 r_{0}^8 \xi  \Gamma \left(\frac{1}{3}\right) \Gamma \left(-\frac{2}{3}\right)-\frac{1}{\left(\frac{1}{Q^2 \xi }\right)^{7/4}}192 \sqrt[4]{6} \pi ^{7/4} r_{0}^{11/3} \sqrt[3]{96 \pi  \xi  Q^2+r_{0}^4}\nonumber\\&&\Bigg(1728 \sqrt{\pi } \xi \Gamma \left(-\frac{2}{3}\right) \Gamma \left(-\frac{1}{4}\right) \Gamma \left(\frac{19}{12}\right) \sqrt{\frac{1}{\xi  Q^2}}+7 \sqrt{6} \Bigg(2 \Gamma \left(-\frac{2}{3}\right) \Bigg(21 \Gamma \left(\frac{5}{4}\right) \nonumber\\&&\Gamma \left(\frac{1}{12}\right)+83 \Gamma \left(\frac{1}{4}\right) \Gamma \left(\frac{13}{12}\right)\Bigg)-315 \Gamma \left(-\frac{3}{4}\right) \Gamma \left(\frac{1}{3}\right) \Gamma \left(\frac{13}{12}\right)\Bigg)\Bigg)-\nonumber\\&&\frac{1}{\left(\frac{1}{\xi  Q^2}\right)^{3/4}}2 \sqrt[4]{6} \pi ^{3/4} r_{0}^{23/3} \sqrt[3]{96 \pi  \xi  Q^2+r_{0}^4}\nonumber\\&&\Bigg(1728 \sqrt{\pi } \xi  \Gamma \left(-\frac{2}{3}\right) \Gamma \left(-\frac{1}{4}\right) \Gamma \left(\frac{19}{12}\right) \sqrt{\frac{1}{\xi  Q^2}}+7 \sqrt{6} \Bigg(2 \Gamma \left(-\frac{2}{3}\right) \Bigg(21 \Gamma \left(\frac{5}{4}\right)\nonumber\\&& \Gamma \left(\frac{1}{12}\right)+83 \Gamma \left(\frac{1}{4}\right) \Gamma \left(\frac{13}{12}\right)\Bigg)-315 \Gamma \left(-\frac{3}{4}\right) \Gamma \left(\frac{1}{3}\right) \Gamma \left(\frac{13}{12}\right)\Bigg)\Bigg)\nonumber\\&&+144 (3 \pi )^{2/3} r_{0}^6 \xi  \left(r_{0}^4+352 \pi  Q^2 \xi \right) \nonumber\\&&\sqrt[3]{\frac{2 r_{0}^4}{Q^2 \xi }+192 \pi } \Gamma \left(\frac{1}{3}\right) \Gamma \left(-\frac{2}{3}\right) \, _2F_1\left(\frac{1}{3},\frac{7}{12};\frac{19}{12};-\frac{r_{0}^4}{96 \pi  Q^2 \xi }\right)\Bigg)\Bigg)\Bigg]\,,
\end{eqnarray}
This result describes the total gravitational mass of the wormhole as measured by a distant observer at spatial infinity. Figure~\ref{fm} illustrates how the ADM mass varies with the electric charge $Q$ for different values of the non-minimal coupling $\xi$, highlighting the combined effects of charge and quantum corrections. The inclusion of the Euler-Heisenberg correction term, parametrized by $\alpha$, plays a crucial role in modifying the ADM mass. Within the Einstein-Euler-Heisenberg (EEH) framework, the ADM mass encapsulates contributions from geometric, electromagnetic, and quantum (nonlinear electrodynamic) sources. As $\alpha$ increases, the quantum correction term becomes increasingly negative, thereby reducing the overall mass. This suggests that quantum effects, encoded in the nonlinear electrodynamics, exert a repulsive or stabilizing influence on the wormhole structure. Consequently, the quantum vacuum fluctuations may effectively reduce the requirement for exotic matter and enhance the wormhole’s stability and physical viability.

\section{Energy Conditions}\label{sec5}
On the basis of the preceding results, we now examine the energy conditions by analyzing their behavior through regional plots. The weak energy condition (WEC) is defined by the inequality
\begin{equation}
T_{\mu \nu }U^{\mu }U^{\nu } \geq 0,
\end{equation}
which simplifies to
\begin{equation}
\rho(r) + p_r(r) \geq 0,
\end{equation}
where $T_{\mu \nu}$ is the energy-momentum tensor and $U^{\mu}$ is a timelike vector. This condition implies that the local energy density measured by any observer is nonnegative. Similarly, the null energy condition (NEC) is given by
\begin{equation}
T_{\mu \nu }k^{\mu }k^{\nu } \geq 0,
\end{equation}
for any null vector $k^{\mu}$, leading to the same inequality:
\begin{equation}
\rho(r) + p_r(r) \geq 0.
\end{equation}

\begin{figure}
\includegraphics[width=8cm]{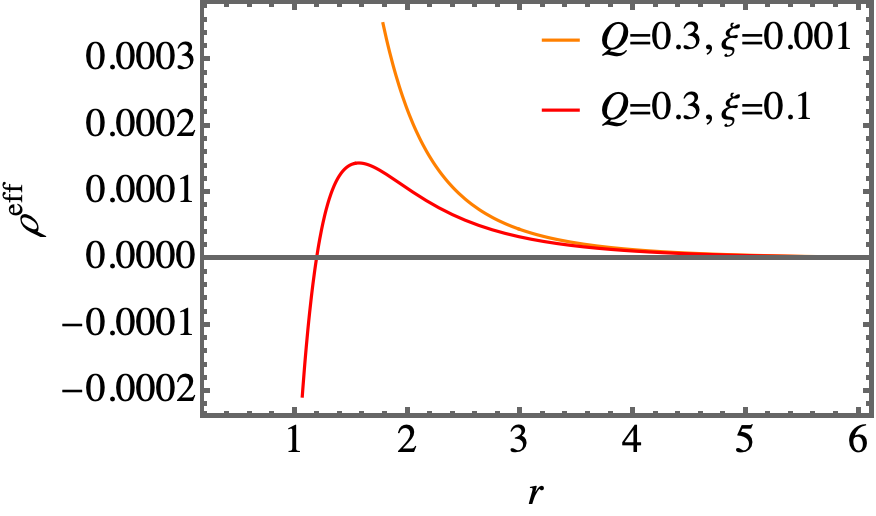}
\includegraphics[width=8cm]{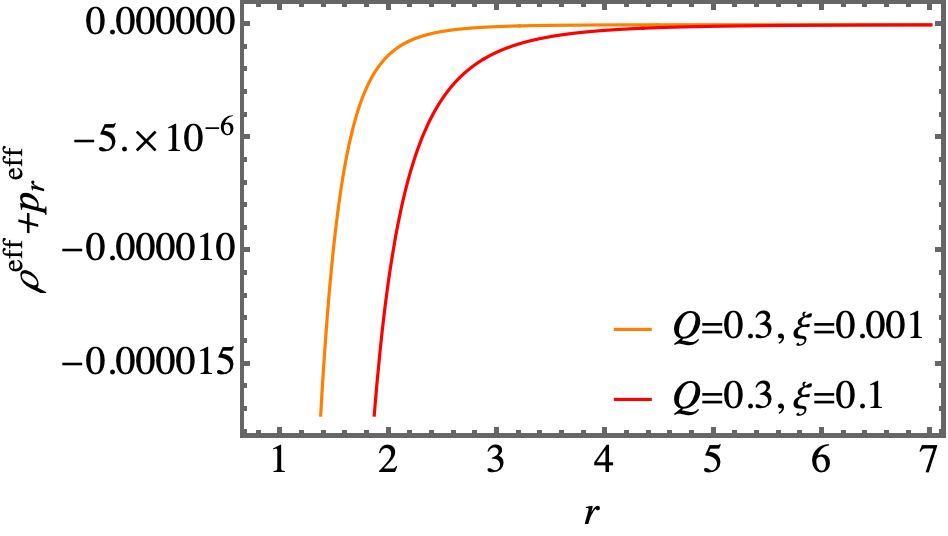}
\includegraphics[width=8cm]{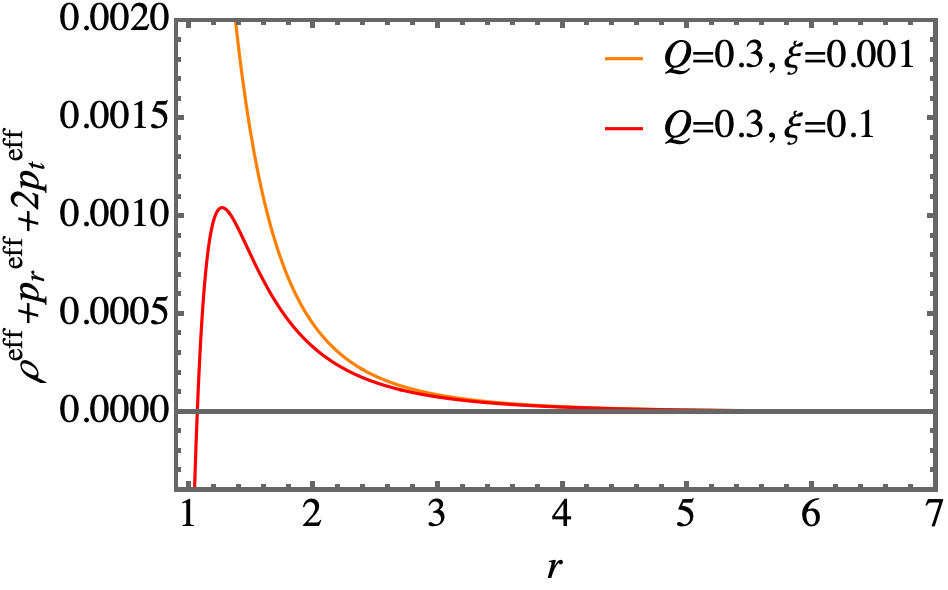}
\includegraphics[width=8cm]{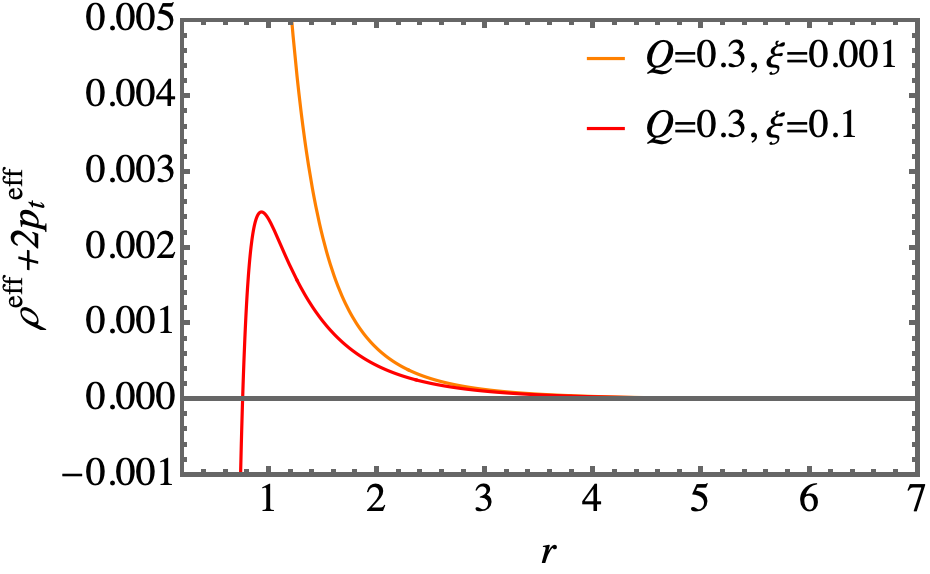}
\caption{The variation of $\rho^{\rm eff},\,\rho^{\rm eff} + p^{\rm eff}_r,\,\rho^{\rm eff} + p^{\rm eff}_r+2p^{\rm eff}_{t}$ and $\rho^{\rm eff} +2p^{\rm eff}_{t}$ as a function of \(r\) using \(\Phi = cont.\). Here we have used $r_0=1,\,Q=0.3$, and $\xi=0.001,\,0.1$.}\label{wec}
\end{figure}

The strong energy condition (SEC) imposes the following constraints:
\begin{equation}
\rho(r) + 2p_t(r) \geq 0, \quad \text{and} \quad \rho(r) + p_r(r) + 2p_t(r) \geq 0.
\end{equation}
From the plotted results, it is clear that both the WEC and NEC are violated at the wormhole throat ($r = r_0$) as illustrated in Fig.~\ref{wec}, while the SEC holds at the same location, as shown in Fig.~\ref{wec}. These outcomes are sensitive to the non-minimal coupling $\xi$. Numerical analysis at the throat $r_0 = 1$ confirms the following:
\begin{equation}
(\rho + p_r)\big|_{r_0=1} < 0, \quad \text{and} \quad (\rho + p_r + 2p_t)\big|_{r_0=1} > 0,
\end{equation}
indicating slight but significant deviations.
Figure~\ref{wec} provides a detailed depiction of the effective energy conditions for the non-minimal Einstein--Yang--Mills (EYM) wormhole, showcasing the profiles of \( \rho_{\text{eff}} \), \( \rho_{\text{eff}} + p^{\text{eff}}_r \), \( \rho_{\text{eff}} + 2p^{\text{eff}}_t \), and \( \rho_{\text{eff}} + p^{\text{eff}}_r + 2p^{\text{eff}}_t \) as functions of the radial coordinate \( r \), under a constant redshift function \( \Phi = \text{const.} \). The figure considers a throat radius \( r_0 = 1 \), charge \( q = 0.3 \), and two values of the non-minimal coupling parameter \( \xi = 0.001 \) and \( \xi = 0.1 \), allowing for an analysis of how non-minimal coupling influences energy condition violations.

The results clearly show that both the WEC and NEC are violated at the wormhole throat \( r = r_0 \), as indicated by the negativity of the quantity \( \rho_{\text{eff}} + p^{\text{eff}}_r \). This is a common feature of traversable wormhole solutions, where exotic matter is traditionally required. In this model, however, the violation emerges naturally from the effective stress-energy tensor induced by the non-minimal coupling between gravity and the Yang-Mills field. Notably, increasing the coupling constant \( \xi \) appears to slightly reduce the extent of violation, hinting that stronger non-minimal interactions can help moderate the exotic matter requirement.

In contrast, the SEC) is satisfied at the throat, especially in the combination \( \rho_{\text{eff}} + p^{\text{eff}}_r + 2p^{\text{eff}}_t \), regardless of the chosen value of \( \xi \). This partial fulfillment of classical energy conditions is significant. It suggests that while some degree of exoticity remains (due to WEC and NEC violations), the quantum and gauge-induced contributions in the non-minimal EYM framework may replace or reduce the need for additional exotic matter sources. This behavior aligns with insights from quantum field theory, where vacuum fluctuations naturally lead to violations of classical energy bounds. Additionally, the positivity of the term \( \rho_{\text{eff}} + 2p^{\text{eff}}_t \), which influences the transverse pressure balance, indicates a stabilizing effect in the angular directions. This contributes to maintaining the wormhole's structural integrity against lateral perturbations. Additionally, Figure~\ref{wec} underscores that non-minimal EYM wormholes can support traversable geometries with localized and softened violations of energy conditions. The dependence on the parameter \( \xi \) supports the idea that suitable tuning of the non-minimal coupling allows for the realization of physically viable wormhole solutions, even in the presence of a constant redshift function.

\section{Light Deflection by non-minimal EYM wormholes}\label{sec6}

\begin{figure}
\includegraphics[width=8 cm]{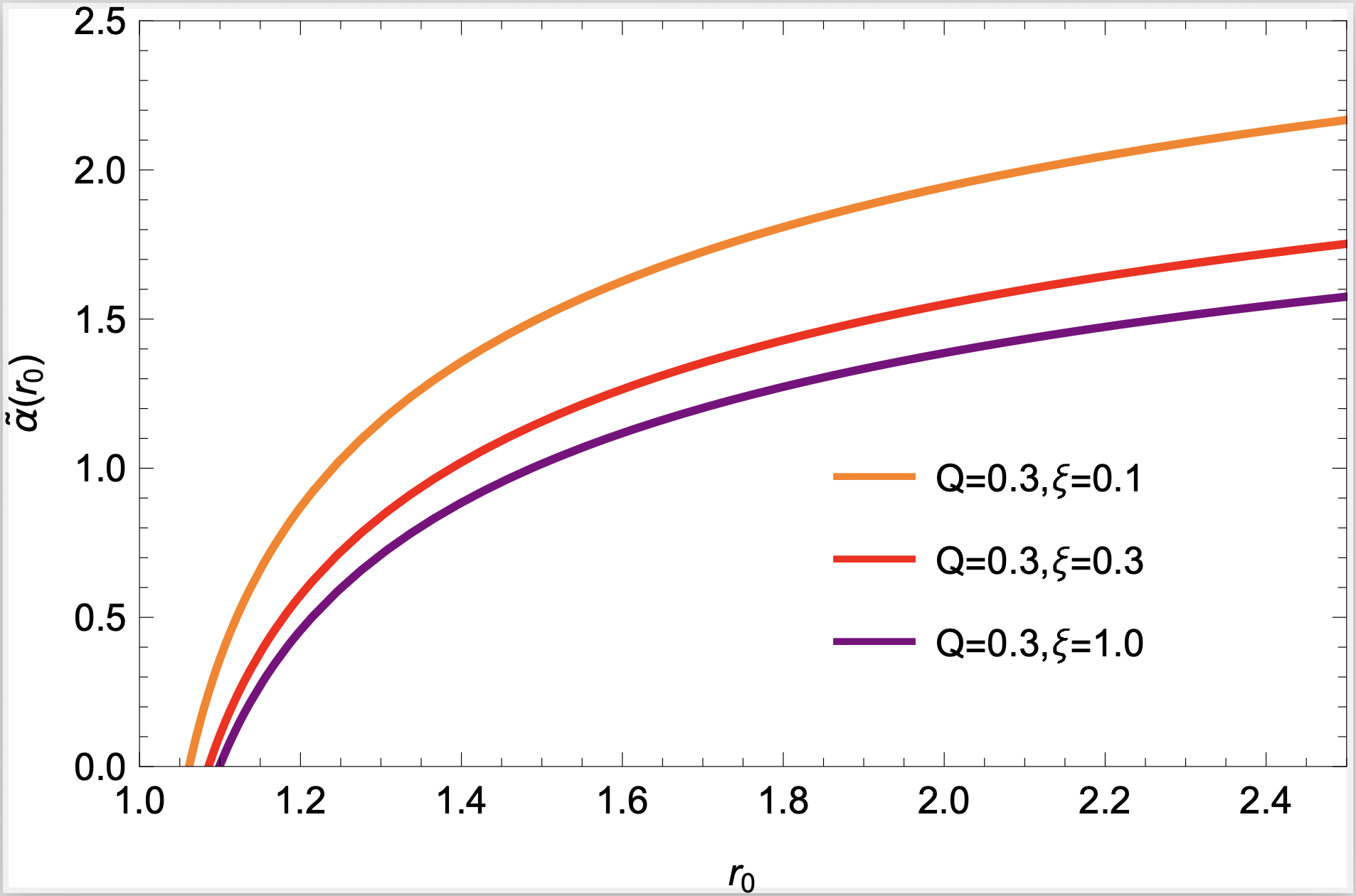}
\includegraphics[width=8 cm]{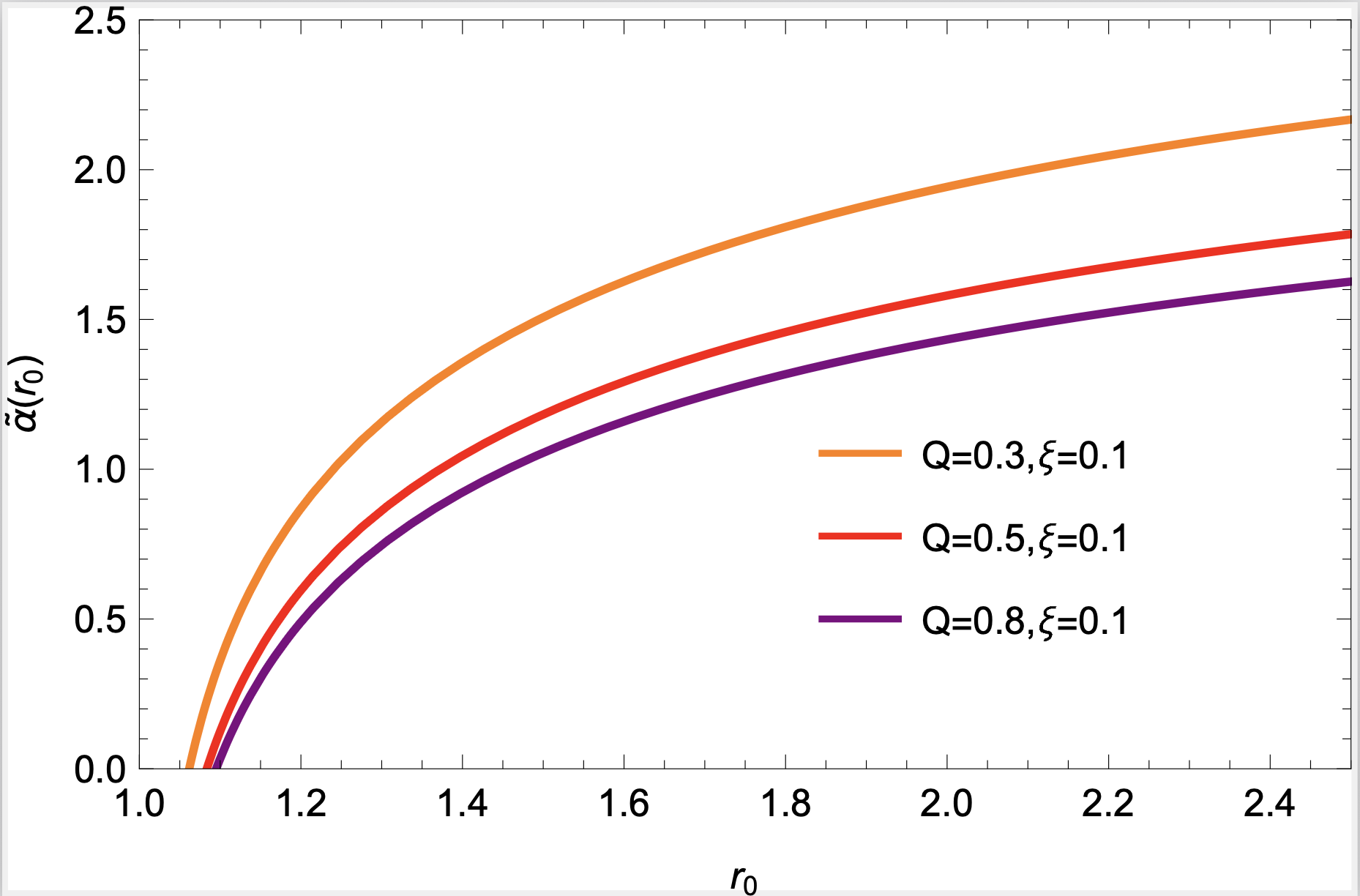}
\caption{Deflection angle of light as a function of $r$ for various values of the coupling constant $\xi$, plotted for $Q = 0.3$ (left) and for various values of $Q$ for $\xi = 0.1$ (right).}\label{angle}
\end{figure}

To analyze the bending of light rays following null geodesics, we begin by considering a general static and spherically symmetric spacetime metric, as outlined in \cite{misner1973freeman,schutz2022first}:

\begin{eqnarray}\label{defeq1}
    d s^2=-A(r) d t^2+B(r) d r^2+C(r) d \Omega^2 .
\end{eqnarray}

The equation governing geodesic motion, which connects the momentum one-forms of a test particle in free fall to the underlying spacetime geometry, can be expressed as \cite{schutz2022first}:

\begin{eqnarray}\label{defeq2}
    \frac{d p_\beta}{d \lambda}=\frac{1}{2} g_{v \alpha, \beta} p^v p^\alpha,
\end{eqnarray}
where, $\lambda$ denotes the affine parameter. It is evident that when the metric components $g_{\alpha \nu}$ do not depend on a particular coordinate $x^\beta$, the corresponding momentum component $p_\beta$ remains conserved along the geodesic. Specifically, by restricting our analysis to the equatorial plane by choosing $\theta = \pi/2$, the metric coefficients $g_{\alpha \beta}$ become independent of the coordinates $t$, $\theta$, and $\phi$ in Eq.~(\ref{defeq2}). This implies the existence of Killing vector fields of the form $\delta_\alpha^\mu \partial_\nu$, where the index $\alpha$ corresponds to a cyclic coordinate. Consequently, the conserved quantities associated with the coordinates $t$ and $\phi$ can be identified as $p_t$ and $p_\phi$, respectively:

\begin{eqnarray}\label{defeq3}
    p_t=-E, \quad p_\phi=L,
\end{eqnarray}

In this context, $E$ and $L$ represent the photon's conserved energy and angular momentum, respectively. Therefore, we can write:

\begin{eqnarray}
    & p_t=\dot{t}=g^{t v} p_v=\frac{E}{A(r)}, \nonumber\\
& p_\phi=\dot{\phi}=g^{\phi v} p_v=\frac{L}{C(r)},\label{defeq4}
\end{eqnarray}

here, an overdot denotes differentiation with respect to the affine parameter $\lambda$. Furthermore, the expression for the radial null geodesic can be straightforwardly derived as:

\begin{eqnarray}\label{defeq5}
    \dot{r}^2=\frac{1}{B(r)}\left(\frac{E^2}{A(r)}-\frac{L^2}{C(r)}\right).
\end{eqnarray}

Nevertheless, the photon's trajectory can also be expressed in terms of the impact parameter $\mu = \frac{L}{E}$, leading to the following form:

\begin{eqnarray}\label{defeq6}
    \left(\frac{d r}{d \phi}\right)^2=\frac{C(r)^2}{\mu^2 B(r)}\left[\frac{1}{A(r)}-\frac{\mu^2}{C(r)}\right] .
\end{eqnarray}

The deflection angle of a photon can be determined by analyzing a scenario where the photon originates from a source with radius $r_s$, which influences the surrounding spacetime geometry. For the photon to reach the surface, there must exist a solution $r_0$ such that $r_0 > r_s$, and the condition $\dot{r}^2 = 0$ holds. The quantity $r_0$ represents the distance of closest approach or the turning point of the trajectory. Under these conditions, the impact parameter takes the form:

\begin{eqnarray}\label{defeq7}
    \mu=\frac{L}{E}= \pm \sqrt{\frac{C\left(r_0\right)}{A\left(r_0\right)}},
\end{eqnarray}

In the weak gravitational field limit, it is evident that $\mu \approx \sqrt{C(r_0)}$. Suppose a photon originates from a distant point with polar coordinates approaching $\lim_{r \rightarrow \infty}\left(r, -\frac{\pi}{2} - \frac{\alpha}{2}\right)$, travels through the turning point at $\left(r_0, 0\right)$, and continues to $\lim_{r \rightarrow \infty}\left(r, \frac{\pi}{2} + \frac{\alpha}{2}\right)$. In that case, the angle of deflection, denoted by $\alpha$, characterizes the total bending of the photon trajectory and depends on the closest approach distance $r_0$ \cite{bhattacharya2010bending}. This deflection angle can be evaluated from Eq.~(\ref{defeq6}) as:

\begin{eqnarray}\label{defeq8}
   \overset{\sim}{\alpha}(r_0)=-\pi+2 \int_{r_0}^{\infty} \frac{\sqrt{B(r)} d r}{\sqrt{C(r)}\left[\left(\frac{A\left(r_0\right)}{A(r)}\right)\left(\frac{C(r)}{C\left(r_0\right)}\right)-1\right]^{1 / 2}}.
\end{eqnarray}

For the specific choices of the metric components characterizing the wormhole geometry, the expression for the deflection angle simplifies to:

\begin{eqnarray}\label{defeq9}
    \overset{\sim}{\alpha}(r_0)=-\pi+2 \int_{r_0}^{\infty} \frac{d r}{r\left[\left(1-\frac{b(r)}{r}\right)\left(\frac{r^2}{r_0^2}-1\right)\right]^{1 / 2}}.
\end{eqnarray}
In evaluating Eq.~(\ref{defeq9}), we substitute the explicit form of the shape function 
$b(r)$ given in Eq.~(\ref{brf}) and perform the integration over the radial coordinate 
from $r=r_{0}$ to spatial infinity. Since the integrand develops an integrable 
square-root singularity at the lower bound, the integration was carried out 
using high-precision adaptive quadrature, ensuring numerical stability near 
the throat. The upper limit was replaced by a sufficiently large cutoff 
$r_{\max}$, chosen so that the remaining contribution from $[r_{\max},\infty)$ 
is negligible. Here, $r_{\text{th}}$ denotes the wormhole throat radius defined 
by $b(r_{\text{th}})=r_{\text{th}}$ with $b'(r_{\text{th}})<1$, while $r_{\max}$ 
represents the numerical cutoff radius used to approximate the integration up 
to infinity. In practice, values of $r_{\max}$ several orders larger than 
$r_{0}$ were sufficient to guarantee convergence. The hypergeometric functions 
and cube roots entering $b(r)$ were evaluated on their real branches, and we 
restricted to parameter domains with $r_{0}\geq r_{\text{th}}$ and 
$1-b(r)/r>0$ for all $r\geq r_{0}$, ensuring that the integrand remains real 
along the photon trajectory. Convergence of the results was checked by 
increasing $r_{\max}$ and tightening error tolerances, leading to the 
consistently positive deflection angles.

The deflection angle of photons in the framework of non-minimal Einstein–Yang–Mills gravity can now be computed by numerically integrating the above equation, incorporating the shape function defined in Eq.~(\ref{brf}). The resulting behavior of the deflection angle is illustrated in Fig.~(\ref{angle}). It is important to know that negative deflection angle shows the repulsive nature of gravity wheas the positive deflection angle shows the attractive nature. It can be clearly verified that the deflection angle does not take negative values in this case \cite{mishra2018trajectories}. So, in our case study of non-minimal Einstein–Yang–Mills gravity. deflection angle of light ray exhibit attractive force.

\section{Conclusion}
In this work, we have developed a new class of traversable wormhole solutions within the framework of non-minimal Einstein–Yang–Mills (EYM) gravity, incorporating a purely magnetic SU(2) Yang–Mills field. By adopting a non-zero redshift function and examining the effects of non-minimal coupling, we have investigated the geometric configuration, energy conditions, ADM mass, and light deflection properties of these wormholes.
Our findings reveal that the non-minimal coupling constant ($\xi$) and the Yang–Mills magnetic charge (Q) play central roles in determining the shape and physical viability of the wormhole geometry. For small values of $\xi$, the wormhole satisfies the flare-out condition and remains traversable without invoking explicit exotic matter. However, larger values of $\xi$ and $Q$ lead to deviations from these conditions, suggesting the emergence of more intricate or quantum-modified geometries. The ADM mass calculation shows that quantum corrections, modeled via the Euler–Heisenberg term, contribute to a reduction in the total gravitational mass, for $\xi \gtrsim 0.01$, which may help stabilize the wormhole structure. Moreover, the energy condition analysis demonstrates that while the Null and Weak Energy Conditions are violated at the throat-consistent with traversable wormhole scenarios-the Strong Energy Condition can still be satisfied under certain parameter choices. The effective stress-energy tensor from the non-minimal coupling produces the necessary violations of the NEC and WEC without adding ad hoc exotic matter components. In other words, the exoticity arises naturally from the geometry–gauge coupling rather than from externally prescribed matter sources.

Additionally, our study of gravitational lensing confirms that the deflection angles of light remain positive, indicating an overall attractive gravitational behavior. This reinforces the plausibility of these wormholes as physically viable solutions and suggests that they may produce observable gravitational lensing effects distinguishable from those of black holes. Looking ahead, several promising directions emerge for future research. A natural extension involves analyzing test particle dynamics and the quasi-normal mode spectra associated with non-minimal EYM wormholes, which could offer unique signatures different from those of black holes. Studies of optical observables such as lensing rings and brightness profiles may further assist in distinguishing wormholes from other compact astrophysical objects. Moreover, recent investigations in modified gravity and quantum-corrected spacetimes have highlighted the influence of geodesic motion, thermodynamic behavior, and stability criteria in both black hole and wormhole contexts. Incorporating these insights could deepen our understanding of the physical implications of non-minimal couplings.

The continued development of observational capabilities, particularly through initiatives such as the Event Horizon Telescope (EHT) \cite{EventHorizonTelescope:2019ggy} and the forthcoming Legacy Survey of Space and Time (LSST)\footnote{\url{http://www.lsst.org/lsst}}, holds the potential to test these theoretical predictions. Distinctive features in wormhole shadows, light deflection patterns, polarization behavior, and gravitational wave emissions may serve as powerful probes for identifying wormhole geometries in astrophysical settings. Future work should aim to refine these theoretical models and establish robust connections with observational data, particularly in scenarios involving rotating wormholes, anisotropic matter distributions, and noncommutative or higher-dimensional gravitational frameworks. Such efforts may ultimately contribute to distinguishing wormholes from black holes and offer new insights into the fundamental nature of spacetime.


\bibliography{ref}

\begin{thebibliography}{61}
\expandafter\ifx\csname natexlab\endcsname\relax\def\natexlab#1{#1}\fi
\expandafter\ifx\csname bibnamefont\endcsname\relax
  \def\bibnamefont#1{#1}\fi
\expandafter\ifx\csname bibfnamefont\endcsname\relax
  \def\bibfnamefont#1{#1}\fi
\expandafter\ifx\csname citenamefont\endcsname\relax
  \def\citenamefont#1{#1}\fi
\expandafter\ifx\csname url\endcsname\relax
  \def\url#1{\texttt{#1}}\fi
\expandafter\ifx\csname urlprefix\endcsname\relax\def\urlprefix{URL }\fi
\providecommand{\bibinfo}[2]{#2}
\providecommand{\eprint}[2][]{\url{#2}}

\bibitem[{\citenamefont{Visser}(1995)}]{visser1995lorentzian}
\bibinfo{author}{\bibfnamefont{M.}~\bibnamefont{Visser}}, \bibinfo{journal}{Woodbury}  (\bibinfo{year}{1995}).

\bibitem[{\citenamefont{Morris and Thorne}(1988{\natexlab{a}})}]{morris1988wormholes}
\bibinfo{author}{\bibfnamefont{M.~S.} \bibnamefont{Morris}} \bibnamefont{and} \bibinfo{author}{\bibfnamefont{K.~S.} \bibnamefont{Thorne}}, \bibinfo{journal}{American Journal of Physics} \textbf{\bibinfo{volume}{56}}, \bibinfo{pages}{395} (\bibinfo{year}{1988}{\natexlab{a}}).

\bibitem[{\citenamefont{Morris et~al.}(1988)\citenamefont{Morris, Thorne, and Yurtsever}}]{morris1988wormholes1}
\bibinfo{author}{\bibfnamefont{M.~S.} \bibnamefont{Morris}}, \bibinfo{author}{\bibfnamefont{K.~S.} \bibnamefont{Thorne}}, \bibnamefont{and} \bibinfo{author}{\bibfnamefont{U.}~\bibnamefont{Yurtsever}}, \bibinfo{journal}{Physical Review Letters} \textbf{\bibinfo{volume}{61}}, \bibinfo{pages}{1446} (\bibinfo{year}{1988}).

\bibitem[{\citenamefont{Visser et~al.}(2003)\citenamefont{Visser, Kar, and Dadhich}}]{visser2003traversable}
\bibinfo{author}{\bibfnamefont{M.}~\bibnamefont{Visser}}, \bibinfo{author}{\bibfnamefont{S.}~\bibnamefont{Kar}}, \bibnamefont{and} \bibinfo{author}{\bibfnamefont{N.}~\bibnamefont{Dadhich}}, \bibinfo{journal}{Physical review letters} \textbf{\bibinfo{volume}{90}}, \bibinfo{pages}{201102} (\bibinfo{year}{2003}).

\bibitem[{\citenamefont{Lobo}(2005)}]{lobo2005phantom}
\bibinfo{author}{\bibfnamefont{F.~S.} \bibnamefont{Lobo}}, \bibinfo{journal}{Physical Review D—Particles, Fields, Gravitation, and Cosmology} \textbf{\bibinfo{volume}{71}}, \bibinfo{pages}{084011} (\bibinfo{year}{2005}).

\bibitem[{\citenamefont{Bl{\'a}zquez-Salcedo et~al.}(2022)\citenamefont{Bl{\'a}zquez-Salcedo, Knoll, and Radu}}]{blazquez2022einstein}
\bibinfo{author}{\bibfnamefont{J.~L.} \bibnamefont{Bl{\'a}zquez-Salcedo}}, \bibinfo{author}{\bibfnamefont{C.}~\bibnamefont{Knoll}}, \bibnamefont{and} \bibinfo{author}{\bibfnamefont{E.}~\bibnamefont{Radu}}, \bibinfo{journal}{The European Physical Journal C} \textbf{\bibinfo{volume}{82}}, \bibinfo{pages}{533} (\bibinfo{year}{2022}).

\bibitem[{\citenamefont{Capozziello et~al.}(2021)\citenamefont{Capozziello, Luongo, and Mauro}}]{capozziello2021traversable}
\bibinfo{author}{\bibfnamefont{S.}~\bibnamefont{Capozziello}}, \bibinfo{author}{\bibfnamefont{O.}~\bibnamefont{Luongo}}, \bibnamefont{and} \bibinfo{author}{\bibfnamefont{L.}~\bibnamefont{Mauro}}, \bibinfo{journal}{The European Physical Journal Plus} \textbf{\bibinfo{volume}{136}}, \bibinfo{pages}{1} (\bibinfo{year}{2021}).

\bibitem[{\citenamefont{Bl{\'a}zquez-Salcedo et~al.}(2021{\natexlab{a}})\citenamefont{Bl{\'a}zquez-Salcedo, Knoll, and Radu}}]{blazquez2021traversable}
\bibinfo{author}{\bibfnamefont{J.~L.} \bibnamefont{Bl{\'a}zquez-Salcedo}}, \bibinfo{author}{\bibfnamefont{C.}~\bibnamefont{Knoll}}, \bibnamefont{and} \bibinfo{author}{\bibfnamefont{E.}~\bibnamefont{Radu}}, \bibinfo{journal}{Physical Review Letters} \textbf{\bibinfo{volume}{126}}, \bibinfo{pages}{101102} (\bibinfo{year}{2021}{\natexlab{a}}).

\bibitem[{\citenamefont{Berry et~al.}(2020)\citenamefont{Berry, Lobo, Simpson, and Visser}}]{berry2020thin}
\bibinfo{author}{\bibfnamefont{T.}~\bibnamefont{Berry}}, \bibinfo{author}{\bibfnamefont{F.~S.} \bibnamefont{Lobo}}, \bibinfo{author}{\bibfnamefont{A.}~\bibnamefont{Simpson}}, \bibnamefont{and} \bibinfo{author}{\bibfnamefont{M.}~\bibnamefont{Visser}}, \bibinfo{journal}{Physical Review D} \textbf{\bibinfo{volume}{102}}, \bibinfo{pages}{064054} (\bibinfo{year}{2020}).

\bibitem[{\citenamefont{Maldacena and Milekhin}(2021)}]{maldacena2021humanly}
\bibinfo{author}{\bibfnamefont{J.}~\bibnamefont{Maldacena}} \bibnamefont{and} \bibinfo{author}{\bibfnamefont{A.}~\bibnamefont{Milekhin}}, \bibinfo{journal}{Physical Review D} \textbf{\bibinfo{volume}{103}}, \bibinfo{pages}{066007} (\bibinfo{year}{2021}).

\bibitem[{\citenamefont{Wielgus et~al.}(2020)\citenamefont{Wielgus, Hor{\'a}k, Vincent, and Abramowicz}}]{wielgus2020reflection}
\bibinfo{author}{\bibfnamefont{M.}~\bibnamefont{Wielgus}}, \bibinfo{author}{\bibfnamefont{J.}~\bibnamefont{Hor{\'a}k}}, \bibinfo{author}{\bibfnamefont{F.}~\bibnamefont{Vincent}}, \bibnamefont{and} \bibinfo{author}{\bibfnamefont{M.}~\bibnamefont{Abramowicz}}, \bibinfo{journal}{Physical Review D} \textbf{\bibinfo{volume}{102}}, \bibinfo{pages}{084044} (\bibinfo{year}{2020}).

\bibitem[{\citenamefont{Bl{\'a}zquez-Salcedo et~al.}(2021{\natexlab{b}})\citenamefont{Bl{\'a}zquez-Salcedo, Chew, Kunz, and Yeom}}]{blazquez2021ellis}
\bibinfo{author}{\bibfnamefont{J.~L.} \bibnamefont{Bl{\'a}zquez-Salcedo}}, \bibinfo{author}{\bibfnamefont{X.~Y.} \bibnamefont{Chew}}, \bibinfo{author}{\bibfnamefont{J.}~\bibnamefont{Kunz}}, \bibnamefont{and} \bibinfo{author}{\bibfnamefont{D.-h.} \bibnamefont{Yeom}}, \bibinfo{journal}{The European Physical Journal C} \textbf{\bibinfo{volume}{81}}, \bibinfo{pages}{1} (\bibinfo{year}{2021}{\natexlab{b}}).

\bibitem[{\citenamefont{Rueda and Contreras}(2025)}]{rueda2025traversable}
\bibinfo{author}{\bibfnamefont{A.}~\bibnamefont{Rueda}} \bibnamefont{and} \bibinfo{author}{\bibfnamefont{E.}~\bibnamefont{Contreras}}, \bibinfo{journal}{Physical Review D} \textbf{\bibinfo{volume}{111}}, \bibinfo{pages}{044019} (\bibinfo{year}{2025}).

\bibitem[{\citenamefont{Xavier et~al.}(2024)\citenamefont{Xavier, Herdeiro, and Crispino}}]{xavier2024traversable}
\bibinfo{author}{\bibfnamefont{S.~V.} \bibnamefont{Xavier}}, \bibinfo{author}{\bibfnamefont{C.~A.} \bibnamefont{Herdeiro}}, \bibnamefont{and} \bibinfo{author}{\bibfnamefont{L.~C.} \bibnamefont{Crispino}}, \bibinfo{journal}{Physical Review D} \textbf{\bibinfo{volume}{109}}, \bibinfo{pages}{124065} (\bibinfo{year}{2024}).

\bibitem[{\citenamefont{Morris and Thorne}(1988{\natexlab{b}})}]{morris1988wormholes2}
\bibinfo{author}{\bibfnamefont{M.~S.} \bibnamefont{Morris}} \bibnamefont{and} \bibinfo{author}{\bibfnamefont{K.~S.} \bibnamefont{Thorne}}, \bibinfo{journal}{American Journal of Physics} \textbf{\bibinfo{volume}{56}}, \bibinfo{pages}{395} (\bibinfo{year}{1988}{\natexlab{b}}).

\bibitem[{\citenamefont{Rahaman et~al.}(2014)\citenamefont{Rahaman, Kuhfittig, Ray, and Islam}}]{rahaman2014possible}
\bibinfo{author}{\bibfnamefont{F.}~\bibnamefont{Rahaman}}, \bibinfo{author}{\bibfnamefont{P.}~\bibnamefont{Kuhfittig}}, \bibinfo{author}{\bibfnamefont{S.}~\bibnamefont{Ray}}, \bibnamefont{and} \bibinfo{author}{\bibfnamefont{N.}~\bibnamefont{Islam}}, \bibinfo{journal}{The European Physical Journal C} \textbf{\bibinfo{volume}{74}}, \bibinfo{pages}{2750} (\bibinfo{year}{2014}).

\bibitem[{\citenamefont{Kuhfittig}(2014)}]{kuhfittig2014gravitational}
\bibinfo{author}{\bibfnamefont{P.~K.} \bibnamefont{Kuhfittig}}, \bibinfo{journal}{The European Physical Journal C} \textbf{\bibinfo{volume}{74}}, \bibinfo{pages}{2818} (\bibinfo{year}{2014}).

\bibitem[{\citenamefont{Lukmanova et~al.}(2016)\citenamefont{Lukmanova, Kulbakova, Izmailov, and Potapov}}]{lukmanova2016gravitational}
\bibinfo{author}{\bibfnamefont{R.}~\bibnamefont{Lukmanova}}, \bibinfo{author}{\bibfnamefont{A.}~\bibnamefont{Kulbakova}}, \bibinfo{author}{\bibfnamefont{R.}~\bibnamefont{Izmailov}}, \bibnamefont{and} \bibinfo{author}{\bibfnamefont{A.~A.} \bibnamefont{Potapov}}, \bibinfo{journal}{International Journal of Theoretical Physics} \textbf{\bibinfo{volume}{55}}, \bibinfo{pages}{4723} (\bibinfo{year}{2016}).

\bibitem[{\citenamefont{Li and Bambi}(2014)}]{li2014distinguishing}
\bibinfo{author}{\bibfnamefont{Z.}~\bibnamefont{Li}} \bibnamefont{and} \bibinfo{author}{\bibfnamefont{C.}~\bibnamefont{Bambi}}, \bibinfo{journal}{Physical Review D} \textbf{\bibinfo{volume}{90}}, \bibinfo{pages}{024071} (\bibinfo{year}{2014}).

\bibitem[{\citenamefont{Abe}(2010)}]{abe2010gravitational}
\bibinfo{author}{\bibfnamefont{F.}~\bibnamefont{Abe}}, \bibinfo{journal}{The Astrophysical Journal} \textbf{\bibinfo{volume}{725}}, \bibinfo{pages}{787} (\bibinfo{year}{2010}).

\bibitem[{\citenamefont{Toki et~al.}(2011)\citenamefont{Toki, Kitamura, Asada, and Abe}}]{toki2011astrometric}
\bibinfo{author}{\bibfnamefont{Y.}~\bibnamefont{Toki}}, \bibinfo{author}{\bibfnamefont{T.}~\bibnamefont{Kitamura}}, \bibinfo{author}{\bibfnamefont{H.}~\bibnamefont{Asada}}, \bibnamefont{and} \bibinfo{author}{\bibfnamefont{F.}~\bibnamefont{Abe}}, \bibinfo{journal}{The Astrophysical Journal} \textbf{\bibinfo{volume}{740}}, \bibinfo{pages}{121} (\bibinfo{year}{2011}).

\bibitem[{\citenamefont{Jamil et~al.}(2013)\citenamefont{Jamil, Momeni, and Myrzakulov}}]{jamil2013observational}
\bibinfo{author}{\bibfnamefont{M.}~\bibnamefont{Jamil}}, \bibinfo{author}{\bibfnamefont{D.}~\bibnamefont{Momeni}}, \bibnamefont{and} \bibinfo{author}{\bibfnamefont{R.}~\bibnamefont{Myrzakulov}}, \bibinfo{journal}{The European Physical Journal C} \textbf{\bibinfo{volume}{73}}, \bibinfo{pages}{2347} (\bibinfo{year}{2013}).

\bibitem[{\citenamefont{Visser}(1996)}]{visser1996voting}
\bibinfo{author}{\bibfnamefont{M.}~\bibnamefont{Visser}}, \bibinfo{journal}{Behavior and Social Issues} \textbf{\bibinfo{volume}{6}}, \bibinfo{pages}{23} (\bibinfo{year}{1996}).

\bibitem[{\citenamefont{Garattini}(2019)}]{Garattini:2019ivd}
\bibinfo{author}{\bibfnamefont{R.}~\bibnamefont{Garattini}}, \bibinfo{journal}{Eur. Phys. J. C} \textbf{\bibinfo{volume}{79}}, \bibinfo{pages}{951} (\bibinfo{year}{2019}), \eprint{1907.03623}.

\bibitem[{\citenamefont{Samart et~al.}(2022)\citenamefont{Samart, Tangphati, and Channuie}}]{Samart:2021tvl}
\bibinfo{author}{\bibfnamefont{D.}~\bibnamefont{Samart}}, \bibinfo{author}{\bibfnamefont{T.}~\bibnamefont{Tangphati}}, \bibnamefont{and} \bibinfo{author}{\bibfnamefont{P.}~\bibnamefont{Channuie}}, \bibinfo{journal}{Nucl. Phys. B} \textbf{\bibinfo{volume}{980}}, \bibinfo{pages}{115848} (\bibinfo{year}{2022}), \eprint{2107.11375}.

\bibitem[{\citenamefont{Garattini and Tzikas}(2025)}]{Garattini:2025gfq}
\bibinfo{author}{\bibfnamefont{R.}~\bibnamefont{Garattini}} \bibnamefont{and} \bibinfo{author}{\bibfnamefont{A.~G.} \bibnamefont{Tzikas}}, \bibinfo{journal}{Eur. Phys. J. C} \textbf{\bibinfo{volume}{85}}, \bibinfo{pages}{336} (\bibinfo{year}{2025}), \eprint{2502.19995}.

\bibitem[{\citenamefont{Jusufi et~al.}(2020)\citenamefont{Jusufi, Channuie, and Jamil}}]{Jusufi:2020rpw}
\bibinfo{author}{\bibfnamefont{K.}~\bibnamefont{Jusufi}}, \bibinfo{author}{\bibfnamefont{P.}~\bibnamefont{Channuie}}, \bibnamefont{and} \bibinfo{author}{\bibfnamefont{M.}~\bibnamefont{Jamil}}, \bibinfo{journal}{Eur. Phys. J. C} \textbf{\bibinfo{volume}{80}}, \bibinfo{pages}{127} (\bibinfo{year}{2020}), \eprint{2002.01341}.

\bibitem[{\citenamefont{Garattini}(2020)}]{Garattini:2020kqb}
\bibinfo{author}{\bibfnamefont{R.}~\bibnamefont{Garattini}}, \bibinfo{journal}{Eur. Phys. J. C} \textbf{\bibinfo{volume}{80}}, \bibinfo{pages}{1172} (\bibinfo{year}{2020}), \eprint{2008.05901}.

\bibitem[{\citenamefont{Garattini}(2021)}]{Garattini:2021kca}
\bibinfo{author}{\bibfnamefont{R.}~\bibnamefont{Garattini}}, \bibinfo{journal}{Eur. Phys. J. C} \textbf{\bibinfo{volume}{81}}, \bibinfo{pages}{824} (\bibinfo{year}{2021}), \eprint{2107.09276}.

\bibitem[{\citenamefont{Garattini and Faizal}(2025)}]{Garattini:2024jkr}
\bibinfo{author}{\bibfnamefont{R.}~\bibnamefont{Garattini}} \bibnamefont{and} \bibinfo{author}{\bibfnamefont{M.}~\bibnamefont{Faizal}}, \bibinfo{journal}{JCAP} \textbf{\bibinfo{volume}{01}}, \bibinfo{pages}{081} (\bibinfo{year}{2025}), \eprint{2403.15174}.

\bibitem[{\citenamefont{Sarkar et~al.}(2025)\citenamefont{Sarkar, Sarkar, and Bouzenada}}]{Sarkar:2025iiz}
\bibinfo{author}{\bibfnamefont{N.}~\bibnamefont{Sarkar}}, \bibinfo{author}{\bibfnamefont{S.}~\bibnamefont{Sarkar}}, \bibnamefont{and} \bibinfo{author}{\bibfnamefont{A.}~\bibnamefont{Bouzenada}} (\bibinfo{year}{2025}), \eprint{2509.02631}.

\bibitem[{\citenamefont{Einstein and Rosen}(1935)}]{einstein1935particle}
\bibinfo{author}{\bibfnamefont{A.}~\bibnamefont{Einstein}} \bibnamefont{and} \bibinfo{author}{\bibfnamefont{N.}~\bibnamefont{Rosen}}, \bibinfo{journal}{Physical Review} \textbf{\bibinfo{volume}{48}}, \bibinfo{pages}{73} (\bibinfo{year}{1935}).

\bibitem[{\citenamefont{Kim and Lee}(2001)}]{kim2001exact}
\bibinfo{author}{\bibfnamefont{S.-W.} \bibnamefont{Kim}} \bibnamefont{and} \bibinfo{author}{\bibfnamefont{H.}~\bibnamefont{Lee}}, \bibinfo{journal}{Physical Review D} \textbf{\bibinfo{volume}{63}}, \bibinfo{pages}{064014} (\bibinfo{year}{2001}).

\bibitem[{\citenamefont{Rahaman et~al.}(2012)\citenamefont{Rahaman, Islam, Kuhfittig, and Ray}}]{rahaman2012searching}
\bibinfo{author}{\bibfnamefont{F.}~\bibnamefont{Rahaman}}, \bibinfo{author}{\bibfnamefont{S.}~\bibnamefont{Islam}}, \bibinfo{author}{\bibfnamefont{P.}~\bibnamefont{Kuhfittig}}, \bibnamefont{and} \bibinfo{author}{\bibfnamefont{S.}~\bibnamefont{Ray}}, \bibinfo{journal}{Physical Review D—Particles, Fields, Gravitation, and Cosmology} \textbf{\bibinfo{volume}{86}}, \bibinfo{pages}{106010} (\bibinfo{year}{2012}).

\bibitem[{\citenamefont{Lobo et~al.}(2013)\citenamefont{Lobo, Parsaei, and Riazi}}]{lobo2013new}
\bibinfo{author}{\bibfnamefont{F.~S.} \bibnamefont{Lobo}}, \bibinfo{author}{\bibfnamefont{F.}~\bibnamefont{Parsaei}}, \bibnamefont{and} \bibinfo{author}{\bibfnamefont{N.}~\bibnamefont{Riazi}}, \bibinfo{journal}{Physical Review D—Particles, Fields, Gravitation, and Cosmology} \textbf{\bibinfo{volume}{87}}, \bibinfo{pages}{084030} (\bibinfo{year}{2013}).

\bibitem[{\citenamefont{Capozziello et~al.}(2012)\citenamefont{Capozziello, Harko, Koivisto, Lobo, and Olmo}}]{capozziello2012wormholes}
\bibinfo{author}{\bibfnamefont{S.}~\bibnamefont{Capozziello}}, \bibinfo{author}{\bibfnamefont{T.}~\bibnamefont{Harko}}, \bibinfo{author}{\bibfnamefont{T.~S.} \bibnamefont{Koivisto}}, \bibinfo{author}{\bibfnamefont{F.~S.} \bibnamefont{Lobo}}, \bibnamefont{and} \bibinfo{author}{\bibfnamefont{G.~J.} \bibnamefont{Olmo}}, \bibinfo{journal}{Physical Review D—Particles, Fields, Gravitation, and Cosmology} \textbf{\bibinfo{volume}{86}}, \bibinfo{pages}{127504} (\bibinfo{year}{2012}).

\bibitem[{\citenamefont{Capozziello and Francaviglia}(2008)}]{capozziello2008extended}
\bibinfo{author}{\bibfnamefont{S.}~\bibnamefont{Capozziello}} \bibnamefont{and} \bibinfo{author}{\bibfnamefont{M.}~\bibnamefont{Francaviglia}}, \bibinfo{journal}{General Relativity and Gravitation} \textbf{\bibinfo{volume}{40}}, \bibinfo{pages}{357} (\bibinfo{year}{2008}).

\bibitem[{\citenamefont{Capozziello et~al.}(2005)\citenamefont{Capozziello, Cardone, and Troisi}}]{capozziello2005reconciling}
\bibinfo{author}{\bibfnamefont{S.}~\bibnamefont{Capozziello}}, \bibinfo{author}{\bibfnamefont{V.~F.} \bibnamefont{Cardone}}, \bibnamefont{and} \bibinfo{author}{\bibfnamefont{A.}~\bibnamefont{Troisi}}, \bibinfo{journal}{Physical Review D—Particles, Fields, Gravitation, and Cosmology} \textbf{\bibinfo{volume}{71}}, \bibinfo{pages}{043503} (\bibinfo{year}{2005}).

\bibitem[{\citenamefont{Capozziello et~al.}(2011)\citenamefont{Capozziello, De~Laurentis, Odintsov, and Stabile}}]{capozziello2011hydrostatic}
\bibinfo{author}{\bibfnamefont{S.}~\bibnamefont{Capozziello}}, \bibinfo{author}{\bibfnamefont{M.}~\bibnamefont{De~Laurentis}}, \bibinfo{author}{\bibfnamefont{S.}~\bibnamefont{Odintsov}}, \bibnamefont{and} \bibinfo{author}{\bibfnamefont{A.}~\bibnamefont{Stabile}}, \bibinfo{journal}{Physical Review D—Particles, Fields, Gravitation, and Cosmology} \textbf{\bibinfo{volume}{83}}, \bibinfo{pages}{064004} (\bibinfo{year}{2011}).

\bibitem[{\citenamefont{Kanti et~al.}(2011)\citenamefont{Kanti, Kleihaus, and Kunz}}]{kanti2011wormholes}
\bibinfo{author}{\bibfnamefont{P.}~\bibnamefont{Kanti}}, \bibinfo{author}{\bibfnamefont{B.}~\bibnamefont{Kleihaus}}, \bibnamefont{and} \bibinfo{author}{\bibfnamefont{J.}~\bibnamefont{Kunz}}, \bibinfo{journal}{Physical review letters} \textbf{\bibinfo{volume}{107}}, \bibinfo{pages}{271101} (\bibinfo{year}{2011}).

\bibitem[{\citenamefont{Kanti et~al.}(2012)\citenamefont{Kanti, Kleihaus, and Kunz}}]{kanti2012stable}
\bibinfo{author}{\bibfnamefont{P.}~\bibnamefont{Kanti}}, \bibinfo{author}{\bibfnamefont{B.}~\bibnamefont{Kleihaus}}, \bibnamefont{and} \bibinfo{author}{\bibfnamefont{J.}~\bibnamefont{Kunz}}, \bibinfo{journal}{Physical Review D—Particles, Fields, Gravitation, and Cosmology} \textbf{\bibinfo{volume}{85}}, \bibinfo{pages}{044007} (\bibinfo{year}{2012}).

\bibitem[{\citenamefont{Arellano and Lobo}(2006{\natexlab{a}})}]{arellano2006evolving}
\bibinfo{author}{\bibfnamefont{A.~V.} \bibnamefont{Arellano}} \bibnamefont{and} \bibinfo{author}{\bibfnamefont{F.~S.} \bibnamefont{Lobo}}, \bibinfo{journal}{Classical and Quantum Gravity} \textbf{\bibinfo{volume}{23}}, \bibinfo{pages}{5811} (\bibinfo{year}{2006}{\natexlab{a}}).

\bibitem[{\citenamefont{Arellano and Lobo}(2006{\natexlab{b}})}]{arellano2006non}
\bibinfo{author}{\bibfnamefont{A.~V.} \bibnamefont{Arellano}} \bibnamefont{and} \bibinfo{author}{\bibfnamefont{F.~S.} \bibnamefont{Lobo}}, \bibinfo{journal}{Classical and Quantum Gravity} \textbf{\bibinfo{volume}{23}}, \bibinfo{pages}{7229} (\bibinfo{year}{2006}{\natexlab{b}}).

\bibitem[{\citenamefont{Faraoni et~al.}(1998)\citenamefont{Faraoni, Gunzig, and Nardone}}]{faraoni1998conformal}
\bibinfo{author}{\bibfnamefont{V.}~\bibnamefont{Faraoni}}, \bibinfo{author}{\bibfnamefont{E.}~\bibnamefont{Gunzig}}, \bibnamefont{and} \bibinfo{author}{\bibfnamefont{P.}~\bibnamefont{Nardone}}, \bibinfo{journal}{arXiv preprint gr-qc/9811047}  (\bibinfo{year}{1998}).

\bibitem[{\citenamefont{Hehl and Obukhov}(2001)}]{hehl2001does}
\bibinfo{author}{\bibfnamefont{F.~W.} \bibnamefont{Hehl}} \bibnamefont{and} \bibinfo{author}{\bibfnamefont{Y.~N.} \bibnamefont{Obukhov}}, in \emph{\bibinfo{booktitle}{Gyros, Clocks, Interferometers...: Testing Relativistic Graviy in Space}} (\bibinfo{publisher}{Springer}, \bibinfo{year}{2001}), pp. \bibinfo{pages}{479--504}.

\bibitem[{\citenamefont{Balakin and Lemos}(2005)}]{balakin2005non}
\bibinfo{author}{\bibfnamefont{A.~B.} \bibnamefont{Balakin}} \bibnamefont{and} \bibinfo{author}{\bibfnamefont{J.~P.} \bibnamefont{Lemos}}, \bibinfo{journal}{Classical and Quantum Gravity} \textbf{\bibinfo{volume}{22}}, \bibinfo{pages}{1867} (\bibinfo{year}{2005}).

\bibitem[{\citenamefont{Channuie et~al.}(2025)\citenamefont{Channuie, Ditta, Kaewkhao, and {\"O}vg{\"u}n}}]{Channuie:2025xlw}
\bibinfo{author}{\bibfnamefont{P.}~\bibnamefont{Channuie}}, \bibinfo{author}{\bibfnamefont{A.}~\bibnamefont{Ditta}}, \bibinfo{author}{\bibfnamefont{N.}~\bibnamefont{Kaewkhao}}, \bibnamefont{and} \bibinfo{author}{\bibfnamefont{A.}~\bibnamefont{{\"O}vg{\"u}n}}, \bibinfo{journal}{Phys. Dark Univ.} \textbf{\bibinfo{volume}{48}}, \bibinfo{pages}{101963} (\bibinfo{year}{2025}), \eprint{2503.23065}.

\bibitem[{\citenamefont{Muller-Hoissen}(1988)}]{muller1988modification}
\bibinfo{author}{\bibfnamefont{F.}~\bibnamefont{Muller-Hoissen}}, \bibinfo{journal}{Classical and Quantum Gravity} \textbf{\bibinfo{volume}{5}}, \bibinfo{pages}{L35} (\bibinfo{year}{1988}).

\bibitem[{\citenamefont{Balakin et~al.}(2007)\citenamefont{Balakin, Sushkov, and Zayats}}]{balakin2007nonminimal}
\bibinfo{author}{\bibfnamefont{A.~B.} \bibnamefont{Balakin}}, \bibinfo{author}{\bibfnamefont{S.~V.} \bibnamefont{Sushkov}}, \bibnamefont{and} \bibinfo{author}{\bibfnamefont{A.~E.} \bibnamefont{Zayats}}, \bibinfo{journal}{Physical Review D—Particles, Fields, Gravitation, and Cosmology} \textbf{\bibinfo{volume}{75}}, \bibinfo{pages}{084042} (\bibinfo{year}{2007}).

\bibitem[{\citenamefont{Balakin and Zayats}(2007)}]{Balakin2005}
\bibinfo{author}{\bibfnamefont{A.~B.} \bibnamefont{Balakin}} \bibnamefont{and} \bibinfo{author}{\bibfnamefont{A.~E.} \bibnamefont{Zayats}}, \bibinfo{journal}{Phys. Lett. B} \textbf{\bibinfo{volume}{644}}, \bibinfo{pages}{294–298} (\bibinfo{year}{2007}).

\bibitem[{\citenamefont{Dyadichev et~al.}(2002)\citenamefont{Dyadichev, Gal’tsov, Zorin, and Zotov}}]{Dyadichev2002}
\bibinfo{author}{\bibfnamefont{V.~V.} \bibnamefont{Dyadichev}}, \bibinfo{author}{\bibfnamefont{D.~V.} \bibnamefont{Gal’tsov}}, \bibinfo{author}{\bibfnamefont{A.~G.} \bibnamefont{Zorin}}, \bibnamefont{and} \bibinfo{author}{\bibfnamefont{M.~Y.} \bibnamefont{Zotov}}, \bibinfo{journal}{Phys. Rev. D} \textbf{\bibinfo{volume}{65}}, \bibinfo{pages}{084007} (\bibinfo{year}{2002}).

\bibitem[{\citenamefont{Bhattacharya and Potapov}(2010{\natexlab{a}})}]{Bhattacharya2010}
\bibinfo{author}{\bibfnamefont{A.}~\bibnamefont{Bhattacharya}} \bibnamefont{and} \bibinfo{author}{\bibfnamefont{A.~A.} \bibnamefont{Potapov}}, \bibinfo{journal}{Mod. Phys. Lett. A} \textbf{\bibinfo{volume}{25}}, \bibinfo{pages}{2399–2409} (\bibinfo{year}{2010}{\natexlab{a}}).

\bibitem[{\citenamefont{Mishra et~al.}(2018)\citenamefont{Mishra, Rahaman, Ray, and Kuhfittig}}]{Mishra2018}
\bibinfo{author}{\bibfnamefont{B.}~\bibnamefont{Mishra}}, \bibinfo{author}{\bibfnamefont{F.}~\bibnamefont{Rahaman}}, \bibinfo{author}{\bibfnamefont{S.}~\bibnamefont{Ray}}, \bibnamefont{and} \bibinfo{author}{\bibfnamefont{P.~K.~F.} \bibnamefont{Kuhfittig}}, \bibinfo{journal}{Eur. Phys. J. C} \textbf{\bibinfo{volume}{78}}, \bibinfo{pages}{113} (\bibinfo{year}{2018}).

\bibitem[{\citenamefont{L{\"u}tf{\"u}o{\u{g}}lu}(2025)}]{Lutfuoglu:2025ljm}
\bibinfo{author}{\bibfnamefont{B.~C.} \bibnamefont{L{\"u}tf{\"u}o{\u{g}}lu}}, \bibinfo{journal}{Eur. Phys. J. C} \textbf{\bibinfo{volume}{85}}, \bibinfo{pages}{630} (\bibinfo{year}{2025}), \eprint{2504.18482}.

\bibitem[{\citenamefont{Balakin et~al.}(2016)\citenamefont{Balakin, Lemos, and Zayats}}]{Balakin:2015gpq}
\bibinfo{author}{\bibfnamefont{A.~B.} \bibnamefont{Balakin}}, \bibinfo{author}{\bibfnamefont{J.~P.~S.} \bibnamefont{Lemos}}, \bibnamefont{and} \bibinfo{author}{\bibfnamefont{A.~E.} \bibnamefont{Zayats}}, \bibinfo{journal}{Phys. Rev. D} \textbf{\bibinfo{volume}{93}}, \bibinfo{pages}{024008} (\bibinfo{year}{2016}), \eprint{1512.02653}.

\bibitem[{\citenamefont{Shaikh}(2018)}]{Shaikh:2018kfv}
\bibinfo{author}{\bibfnamefont{R.}~\bibnamefont{Shaikh}}, \bibinfo{journal}{Phys. Rev. D} \textbf{\bibinfo{volume}{98}}, \bibinfo{pages}{024044} (\bibinfo{year}{2018}), \eprint{1803.11422}.

\bibitem[{\citenamefont{Misner et~al.}(1973)\citenamefont{Misner, Thorne, Wheeler, and Gravitation}}]{misner1973freeman}
\bibinfo{author}{\bibfnamefont{C.~W.} \bibnamefont{Misner}}, \bibinfo{author}{\bibfnamefont{K.~S.} \bibnamefont{Thorne}}, \bibinfo{author}{\bibfnamefont{J.~A.} \bibnamefont{Wheeler}}, \bibnamefont{and} \bibinfo{author}{\bibfnamefont{W.}~\bibnamefont{Gravitation}}, \bibinfo{journal}{San Francisco} \textbf{\bibinfo{volume}{891}} (\bibinfo{year}{1973}).

\bibitem[{\citenamefont{Schutz}(2022)}]{schutz2022first}
\bibinfo{author}{\bibfnamefont{B.}~\bibnamefont{Schutz}}, \emph{\bibinfo{title}{A first course in general relativity}} (\bibinfo{publisher}{Cambridge university press}, \bibinfo{year}{2022}).

\bibitem[{\citenamefont{Bhattacharya and Potapov}(2010{\natexlab{b}})}]{bhattacharya2010bending}
\bibinfo{author}{\bibfnamefont{A.}~\bibnamefont{Bhattacharya}} \bibnamefont{and} \bibinfo{author}{\bibfnamefont{A.~A.} \bibnamefont{Potapov}}, \bibinfo{journal}{Modern Physics Letters A} \textbf{\bibinfo{volume}{25}}, \bibinfo{pages}{2399} (\bibinfo{year}{2010}{\natexlab{b}}).

\bibitem[{\citenamefont{Mishra and Chakraborty}(2018)}]{mishra2018trajectories}
\bibinfo{author}{\bibfnamefont{A.}~\bibnamefont{Mishra}} \bibnamefont{and} \bibinfo{author}{\bibfnamefont{S.}~\bibnamefont{Chakraborty}}, \bibinfo{journal}{The European Physical Journal C} \textbf{\bibinfo{volume}{78}}, \bibinfo{pages}{1} (\bibinfo{year}{2018}).

\bibitem[{\citenamefont{Akiyama et~al.}(2019)}]{EventHorizonTelescope:2019ggy}
\bibinfo{author}{\bibfnamefont{K.}~\bibnamefont{Akiyama}} \bibnamefont{et~al.} (\bibinfo{collaboration}{Event Horizon Telescope}), \bibinfo{journal}{Astrophys. J. Lett.} \textbf{\bibinfo{volume}{875}}, \bibinfo{pages}{L6} (\bibinfo{year}{2019}), \eprint{1906.11243}.

\end{thebibliography}

\end{document}